\documentclass[aip,graphicx]{revtex4-1}

\usepackage{epsfig}
\usepackage{color}
\usepackage{amsmath}
\usepackage{amssymb}
\usepackage{citesort}
\usepackage{booktabs} 
\usepackage{rotating} 
\usepackage{longtable}
\usepackage{dcolumn}
\usepackage{bm}
\usepackage{enumerate}
\usepackage{ulem}
\usepackage{float}
\usepackage{sidecap}

\providecommand{\abs}[1]{\lvert#1\rvert}

\newcommand{\be}{\begin{equation}}
\newcommand{\ee}{\end{equation}}
\newcommand{\ba}{\begin{eqnarray}}
\newcommand{\ea}{\end{eqnarray}}
\newcommand{\nn}{\nonumber \\}

\draft 

\begin{document}


\title{Finding low-energy conformations of lattice protein models by quantum annealing} 

\author{Alejandro Perdomo-Ortiz}
\affiliation{Department  of Chemistry and Chemical Biology, Harvard University, 12 Oxford Street, Cambridge, MA 02138, USA}

\author{Neil Dickson}
\affiliation{D-Wave Systems, Inc., 100-4401 Still Creek Drive, Burnaby, British Columbia V5C 6G9, Canada}

\author{Marshall Drew-Brook}
\affiliation{D-Wave Systems, Inc., 100-4401 Still Creek Drive, Burnaby, British Columbia V5C 6G9, Canada}

\author{Geordie Rose}
\affiliation{D-Wave Systems, Inc., 100-4401 Still Creek Drive, Burnaby, British Columbia V5C 6G9, Canada}

\author{Al\'an Aspuru-Guzik}
\email[Corresponding author's e-mail: ]{aspuru@chemistry.harvard.edu}
\affiliation{Department  of Chemistry and Chemical Biology, Harvard University, 12 Oxford Street, Cambridge, MA 02138, USA}

\date{\today}


\maketitle 


\textbf{Lattice protein folding models are a cornerstone of computational biophysics~\cite{sali_how_1994}. Although these models are a coarse grained representation, they provide useful insight into the energy landscape of natural proteins~\cite{pande_simple_2010,dill_protein_2008,mirny2001,pande_heteropolymer_2000,kolinski1996,shakhnovich_proteins_1994}. Finding low-energy three-dimensional structures is an intractable problem~\cite{Berger1998,Crescenzi1998,hart_robust_1997} even in the simplest model, the Hydrophobic-Polar (HP) model. Exhaustive search of all possible global minima is limited to sequences in the tens of amino acids~\cite{yue_forces_1995}. Description of protein-like properties are more accurately described by generalized models, such as the one proposed by Miyazawa and Jernigan~\cite{miyazawa_residue-residue_1996} (MJ), which explicitly take into account the unique interactions among all 20 amino acids. There is theoretical~\cite{amara_global_1993,finnila_quantum_1994,kadowaki_quantum_1998,Farhi2001,santoro_theory_2002} and experimental~\cite{brooke_quantum_1999} evidence of the advantage of solving classical optimization problems using quantum annealing~\cite{finnila_quantum_1994,kadowaki_quantum_1998,santoro_optimization_2006,das_2008} over its classical analogue (simulated annealing~\cite{kirkpatrick_optimization_1983}). In this report, we present a benchmark implementation of quantum annealing for a biophysical problem (six different experiments up to 81 superconducting quantum bits). Although the cases presented here can be solved in a classical computer, we present the first implementation of lattice protein folding on a quantum device under the Miyazawa-Jernigan model. This paves the way towards studying optimization problems in biophysics and statistical mechanics using quantum devices.}

The search for more efficient optimization algorithms is an important endeavor with prevalence on many disciplines ranging from the social sciences to the physical and natural sciences. Belonging to the latter, the protein folding problem~\cite{dill_protein_2008,mirny2001,kolinski1996,pande_heteropolymer_2000} consists of finding the lowest free-energy configuration or, equivalently, the native structure of a protein given its amino acid sequence. Knowing how proteins fold elucidate their three-dimensional structure-function relationship which is crucial to the understanding of enzymes and for the treatment of misfolded-protein diseases such as Alzheimer's, Huntington's, and Parkinson's disease. Due to the high computational cost of modeling proteins in atomistic detail~\cite{bohannon_distributed_2005,shaw2010}, coarse-grained descriptions of the protein folding problem, such as those found in lattice models, provide valuable insight about the folding mechanisms~\cite{kolinski1996,pande_heteropolymer_2000,mirny2001,pande_simple_2010,Li_PRL2010}.

Harnessing quantum-mechanical effects to speed up the solving of classical optimization problems is at the heart of quantum annealing algorithms (QA)~\cite{finnila_quantum_1994,kadowaki_quantum_1998,santoro_optimization_2006,das_2008}. In QA, quantum mechanical tunneling allows for more efficient exploration of difficult potential energy landscapes such as that of classical spin-glass problems. In our implementation of lattice folding, quantum fluctuations (tunneling) occurs between states representing different model protein conformations or folds.

The theoretical challenge is to efficiently map the hard computational problem of interest (e.g., lattice folding) to a classical spin-glass Hamiltonian: such mapping requiring a polynomial number of quantum bits (qubits) with the size of the problem (protein length) is described elsewhere~\cite{perdomo08}. Here we present a new mapping which, due to its exponential scaling with problem size, is not intended for large instances. The proposed mapping employs very few qubits for small problem instances, making it ideal for this first experimental demonstration and implementation on current quantum devices~\cite{johnson_quantum_2011}. A combination of the existing polynomial mapping~\cite{perdomo08} and more advanced quantum devices would allow for the simulation of much larger instances of lattice folding and other related optimization problems.

Solving arbitrary problem instances requires a programmable quantum device to implement the corresponding classical Hamiltonian. We employ quantum annealing on the programmable device  to obtain low-energy conformations of the protein model. We emphasize that nothing quantum mechanical is implied about the protein or its folding process; rather quantum fluctuations are a tool we use to solve the optimization problem.

The QA protocol performed here is also known as adiabatic quantum computation (AQC)~\cite{farhi2000,Farhi2001}. Of all the quantum-computational models, AQC is perhaps the most naturally suited for studying and solving optimization problems~\cite{Farhi2001,hogg03}. For the experiments presented here, the small finite temperature of the superconducting device is enough to make the process less coherent than the original formulation of AQC, where the theoretical limit of zero temperature and quasi-adiabaticity are usually assumed~\cite{farhi2000,Farhi2001}. As we show in the discussion, numerical simulations including these unavoidable environmental effects accurately reproduce our experimental results.

Experimental implementations of QA or AQC are limited either by the number of qubits available in state-of-the-art quantum devices or by the programmability required to fulfill the problem specification. For example, the first realization of AQC was performed on a three-qubit NMR quantum device~\cite{steffen_experimental_2003} and newer NMR implementations involve four qubit experiments~\cite{Xu2011}. Other experimental realizations of spin systems have been based on measuring bulk magnetization properties of the systems in which there is no control over the individual spins and the couplings among them~\cite{brooke_quantum_1999,wernsdorfer_molecular_2010}. Quantum architectures using superconducting qubits ~\cite{vion2002,You2005,lupascu2007,hofheinz2009,dicarlo2009,neeley_generation_2010,dicarlo_preparation_2010,kaminsky2004} offer promising device scalability while maintaining the ability to control individual qubits and the strength of their interaction couplings. During the preparation of this manuscript, an 84-qubit experimental determination of Ramsey numbers with quantum annealing was performed~\cite{Zhengbing}, underscoring the programmable capabilities of the device for problems with over 80 qubits. In this letter, we present a quantum annealing experimental implementation of lattice protein models with general (Miyazawa-Jernigan~\cite{miyazawa_residue-residue_1996}) interactions among the amino acids. Even though the cases presented here still can be solved on a classical computer by exact enumeration (the six-amino acid problem has only 40 possible configurations), it is remarkable that the device anneals to the ground state of a search space of $2^{81}$ possible computational outcomes. This study provides a proof-of-principle that optimization of biophysical problems such as protein folding can be studied using quantum mechanical devices.



\begin{figure}[H]
\begin{center}
\includegraphics[width=0.8\textwidth]{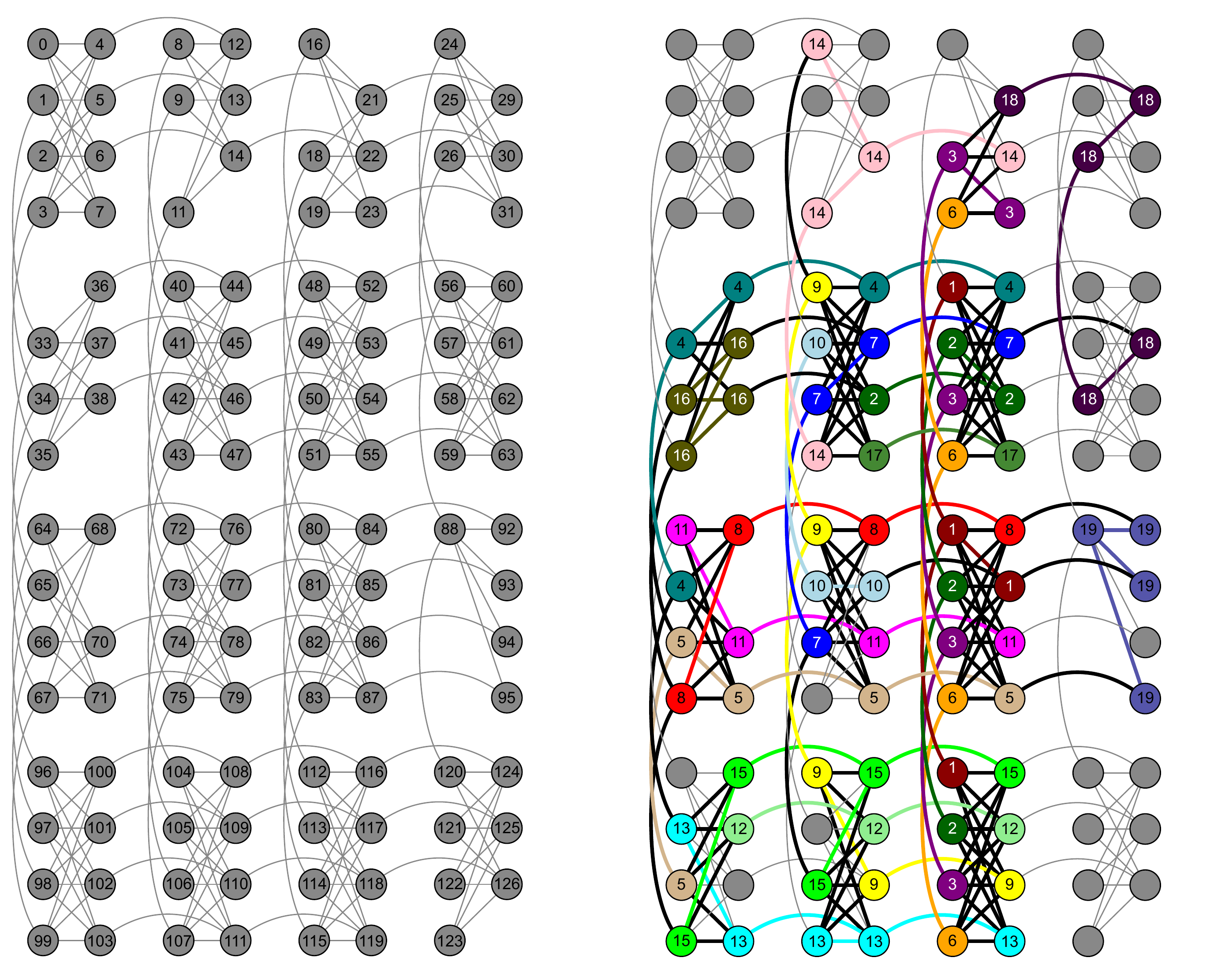}
\end{center}
\caption{The array of superconducting quantum bits is arranged in $4\times 4$ unit cells that consist of 8 quantum bits each. Within a unit cell, each of the 4 qubits in the left-hand partition (LHP) connects to all 4 qubits in the right-hand partition (RHP), and vice versa. A qubit in the LHP (RHP) also connects to the corresponding qubit in the LHP (RHP) of the units cells above and below (to the left and right of) it. (a) Qubits are labeled from 0 to 127 and edges between qubits represent couplers with programmable coupling strengths. Grey qubits indicate the 115 usable qubits, while vacancies indicate qubits under calibration which were not used. The larger experiments (Experiments 1,2, and 4) were performed on this chip, while the three remaining smaller experiments were run on other chips with the same architecture. (b) Embedding and qubit connectivity for Experiment 4, coloring the 81 qubits used in the experiment. Nodes with the same color represent the same logical qubit from the original 19-qubit Ising-like Hamiltonian resulting from the energy function associated with Experiment 4 (see SI material for details). This embedding aims to fulfill the arbitrary connectivity of the Ising expression and allows for the coupling of qubits that are not directly coupled in hardware.}
\label{fig:hardware}
\end{figure}

The quantum hardware employed consists of 16 units of a recently characterized eight-qubit unit cell~\cite{harris2010,johnson_quantum_2011}. Post-fabrication characterization determined that only 115 qubits out of the 128 qubit array can be reliably used for computation (see Fig.~\ref{fig:hardware}). The array of coupled superconducting flux qubits is, effectively, an artificial Ising spin system with programmable spin-spin couplings and transverse magnetic fields. It is designed to solve instances of the following (NP-hard~\cite{Barahona1982}) classical optimization problem: Given a set of local longitudinal fields $\{h_i\}$ and an interaction matrix $\{J_{ij}\}$, find the assignment $\mathbf{s^*} = s^*_1 s^*_2 \cdots s^*_N$, that minimizes the objective function $E(\mathbf{s})$, where,
\begin{equation}\label{eq:QUBO}
E(\mathbf{s})  = \sum_{1 \le i \le N} h_{i} s_i  + \sum_{1 \le i<j\le N} J_{ij} s_{i} s_{j},
\end{equation}
$\abs{h_i} \le 1$, $\abs{J_{ij}} \le 1$, and $s_i \in \{+1,-1\}$.

Finding the optimal $\mathbf{s^*}$ is equivalent to finding the ground state of the corresponding Ising classical Hamiltonian,
\begin{equation}\label{h-ising}
H_{p}  =  \sum^N_{1 \le i \le N} h_{i}\sigma_{i}^{z}  + \sum^N_{1 \le i<j\le N} J_{ij}\sigma_{i}^{z} \sigma_{j}^{z}
\end{equation}
where $\sigma_{i}^{z}$ are Pauli matrices acting on the $i$th spin.

\begin{figure}[H]
 \begin{center}
 \includegraphics[width=0.7\textwidth]{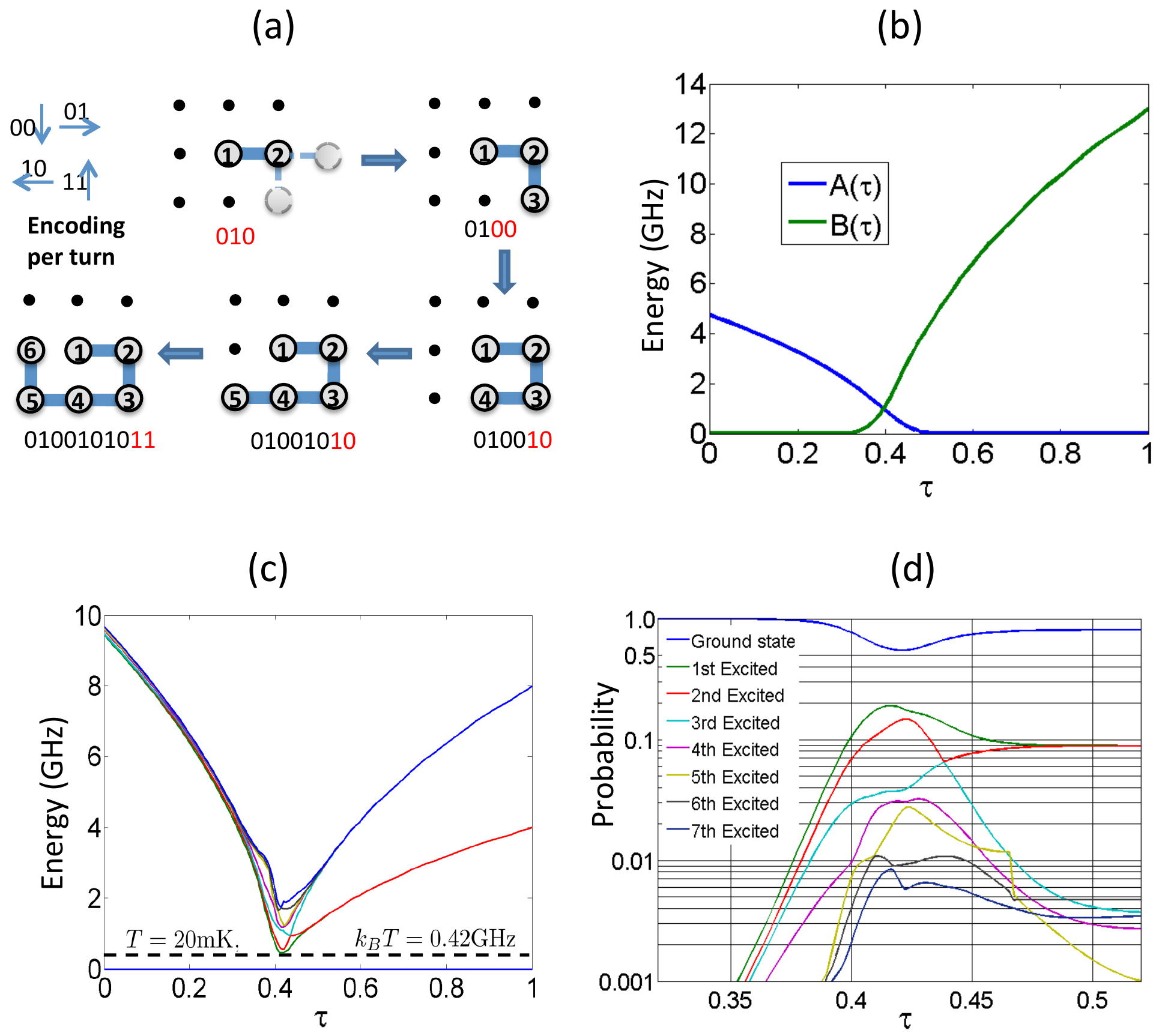}
 \end{center}
 \vspace{-10mm}
 \caption{ \label{fig:encoding} (a) Step-by-step construction of the binary representation of lattice protein. Two qubits per bond are needed and the bond directions are denoted as ``00" (downwards), ``01" (rightwards), ``10" (leftwards), and ``11" (upwards). The example shows one of the possible folds of an arbitrary six-amino-acid sequence.  Any possible $N$-amino-acid fold can be represented by a string of variables $01 0 q_1 q_2 q_3 \cdots q_{\ell-1} q_{\ell}$ with $\ell = 2N-5$. (b)Time-dependence of the $A(\tau)$ and $B(\tau)$ functions, where $\tau=t/t_{run}$ with $t_{run} =148 \mu s$,  (c) time-dependent spectrum obtained through numerical diagonalization, and (d) Bloch-Redfield simulations showing the time-dependent population in the first eight instantaneous eigenstates of the experimentally implemented 8-qubit Hamiltonian (Eq.~\ref{h-AQO}) with $H_p$ from Eq.~18 in the SI material. In panel (c),  for each instantaneous eigenenergy curve we have subtracted the energy of the ground state, effectively plotting the gap of the seven-lowest-excited states with respect to the ground state (represented by the baseline at zero-energy). As a reference, we show the energy  with the device temperature, which is comparable to the minimum gap between the ground and first excited state. In panel (d), populations are ordered in energy from top (ground state) to bottom. Although $\tau = t/t_{run}$ runs from 0 to 1, we show the region where most of the population changes occur. As expected, this is in the proximity of the minimum gap between the ground and first excited state around $\tau \sim 0.4$ [see panel(c)]. }
 \end{figure}

Experimentally, the time-dependent quantum Hamiltonian implemented in the superconducting-qubit array is given by,
\begin{equation}\label{h-AQO}
H(\tau)  = A(\tau) H_b + B(\tau) H_p, \quad \quad \tau= t/t_{run},
\end{equation}
with $H_b  = - \sum_i \sigma^{x}_{i}$ responsible for quantum tunneling among the localized classical states, which correspond to the eigenstates of $H_p$ (the computational basis). The time-dependent functions $A(\tau)$ and $B(\tau)$ are such that $A(0) \gg B(0)$ and $A(1) \ll B(1)$; in Fig.~\ref{fig:encoding}(b), we plot these functions as implemented in the experiment. $t_{run}$ denotes the time elapsed between the preparation of the initial state and the measurement.

QA exploits the adiabatic theorem of quantum mechanics, which states that a quantum system initialized in the ground state of a time-dependent Hamiltonian remains in the instantaneous ground state, as long as it is driven sufficiently slowly. Since the ground state of $H_p$ encodes the solution to the optimization problem, the idea behind QA is to adiabatically prepare this ground state by initializing the quantum system in the easy-to-prepare ground state of $H_b$, which corresponds to a superposition of all $2^N$ states of the computational basis. The system is driven slowly to the problem Hamiltonian, $H(\tau=1) \approx H_p$.  Deviations from the ground-state are expected due to deviations from adiabaticity, as well as thermal noise and imperfections in the implementation of the Hamiltonian.


The first challenge of the experimental implementation is to map the computational problem of interest into the binary quadratic expression (Eq.~\ref{h-ising}), which we outline next.
In lattice folding, the sequence of amino acids defining the protein is viewed as a sequence of beads (amino acids) connected by strings (peptide bonds). This bead chain occupies points on a two- or three-dimensional lattice. A valid configuration is a self-avoiding walk on the lattice and its energy is calculated from the sum of interaction energies between nearest non-bonded neighbors on the lattice.
By the thermodynamic hypothesis of protein folding~\cite{anfinsen_principles_1973}, the global minimum of the free-energy function is conjectured to be the native functional conformation of the protein.

\begin{figure}[H]
\begin{center}
\includegraphics[width=0.95\textwidth]{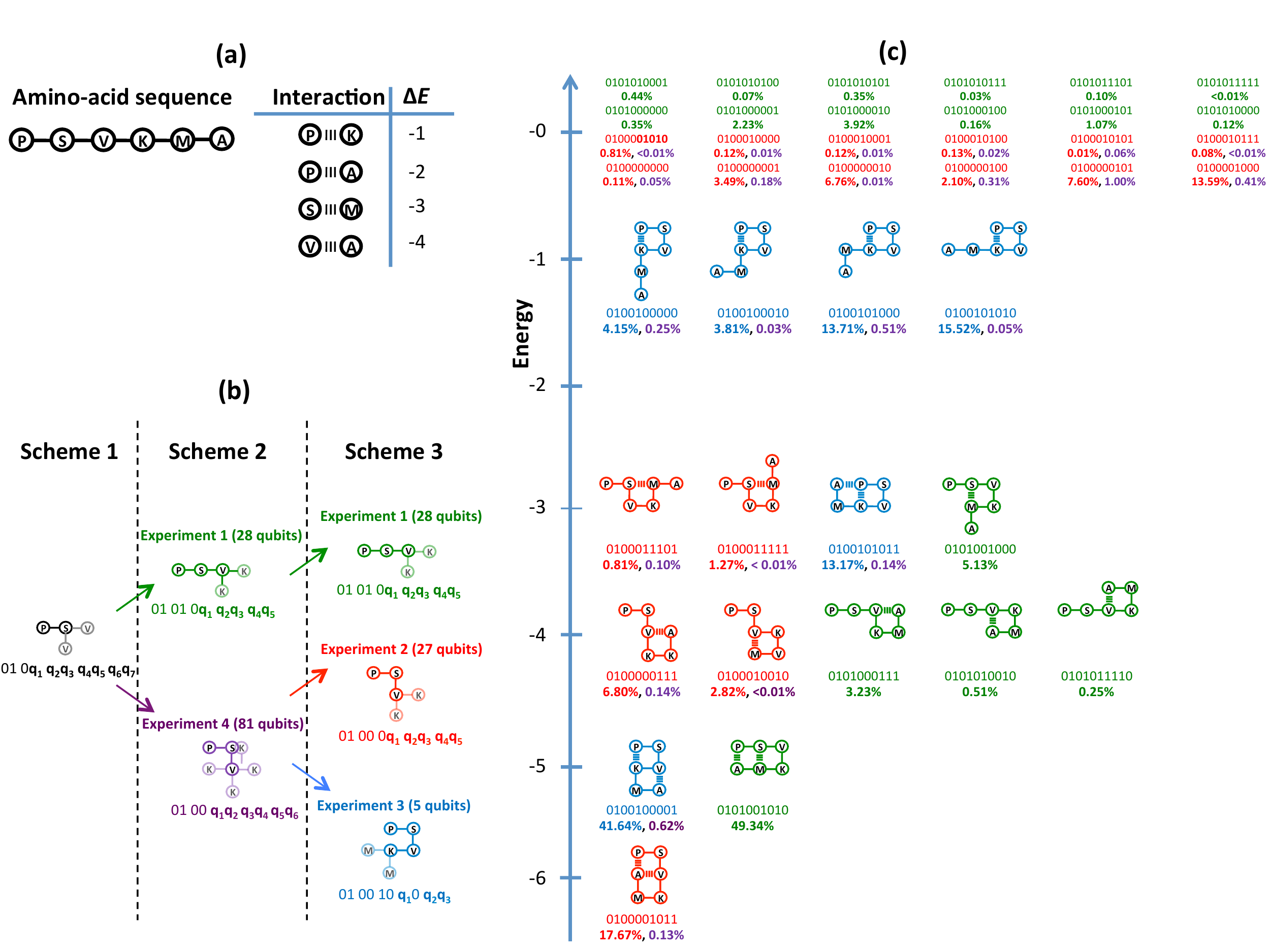}
\end{center}
\caption{(a) Representation of the six-amino acid sequence, Proline-Serine-Valine-Lysine-Methionine-Alanine with its respective one-letter sequence notation, PSVKMA. We use the pairwise nearest-neighbor Miyazawa-Jernigan interaction energies reported in Table 3 of Ref.~\onlinecite{miyazawa_residue-residue_1996}. (b) Divide and conquer approach showing three different schemes which independently solve the six-amino acid sequence PSVKMA on a two-dimensional lattice. We solved the problem under Scheme2 and 3 (Experiments 1 through 4). (c) Energy landscape for the valid conformations of the PSVKMA sequence. Results of the experimentally-measured probability outcomes are given as color-coded percentages according to each of the experimental realizations described in panel (b). Percentages for states with energy greater than zero are 32.70\%, 59.88\%, 8.00\%, and 95.97\% for Experiments 1 through 4, respectively.}
\label{fig:lanscape6AA}
\end{figure}

The hydrophobic-polar (HP) model is one of the simplest possible models for lattice folding~\cite{LAU1989}. In this model, the amino acids are classified into two groups, hydrophobic (H) and polar (P). To describe real protein energy landscapes a more elaborate description needs to be considered, such as the Mijazawa-Jernigan (MJ) model~\cite{miyazawa_residue-residue_1996} which assigns the interaction energies for pairwise interactions among all twenty amino acids. The formulation we used is general enough to take into account arbitrary interaction matrices for lattice models in two and three dimensions~\cite{Perdomo-OrtizTURNmapping2010}. In particular, we solved a MJ model in 2D, the six amino-acid sequence of Proline-Serine-Valine-Lysine-Methionine-Alanine (PSVKMA in the one-letter amino-acid sequence notation). We solved the problem under two different experimental schemes (see Schemes 2 and 3 in Fig.~\ref{fig:lanscape6AA}), each requiring a different number of resources. Solving the problem in one proposed experimental realization (Scheme 1) requires more resources than the number of qubits available (115 qubits) in the device. Scheme 2 and 3 are examples of the divide-and-conquer strategy, in which one partitions the problem in smaller instances and combines the independent set of results, thereby obtaining the same solution for the intractable problem. In the SI section, we complement these four MJ related experiments with two small tetrapeptide instances (effectively HP model instances) for a total of six different problem Hamiltonians. We used the largest of these two instances (an 8 qubit experiment) for direct theoretical simulation of the annealing dynamics of the device. The results from our experiment and the theoretical model, which does not use any adjustable parameters (all are extracted experimentally from the device), are in excellent agreement (see panel (b), Fig.~2 of the SI material).

To represent each of the possible $N$-amino-acid configurations (folds) in the lattice, we encode the direction of each successive bond between amino acids; thus, for every $N$-bead sequence we need to specify $N-1$ turns corresponding to the number of bonds. For the case of a two dimensional lattice, a bond can take any of four possible directions; therefore, two bits per bond are required to uniquely determine a direction. More specifically, if a bond points upwards, we write ``11". If it points downwards, leftwards or rightwards, we write ``00", ``10", or ``01" respectively. Fixing the direction of the first bond reduces the description of any $N$-bead fold to $\ell = 2(N-2)$ binary variables, without loss of generality. As shown in Fig.~\ref{fig:encoding}(a), in the absence of external constraints other than those imposed by the primary amino acid sequence (see SI for an example with external constraints), we can fix the third binary variable to ``0", forcing the third amino acid to go either straight or downward and reducing the number of needed variables to $\ell = 2N-5$. This constraint reduces the solution space by removing conformations which are degenerate due to rotational symmetry. Thus, a particular fold is uniquely defined by,
\begin{equation}\label{eq:q}
\boldsymbol{q} = \underbrace{0 1}_{turn1} \underbrace{0 q_1}_{turn2} \underbrace{q_2 q_3}_{turn3} \cdots  \underbrace{q_{2N-6} q_{2N-5}}_{turn(N-1)}
\end{equation}
An example of this encoding for a six-amino-acid sequence is represented in Fig.~\ref{fig:encoding}(a).

Using this mapping to translate between the amino acid chain in the lattice and the $2(N-1)$ string of bits, we constructed the energy function $E(\boldsymbol{q})$  in which $\boldsymbol{q}$ denotes the remaining $2N-5$ binary variables. Additionally, we penalized folds which exhibit two amino acids on top of each other, to favor self-avoiding walk configurations. The energy penalty chosen for each problem was sufficient to push the energy of invalid folds outside of the energy range of valid configurations (those with $E \le 0$). Finally, we took into account the interaction energy among the different amino acids. A detailed construction of our energy function for the general case of $N$ amino acids with arbitrary interactions is given elsewhere~\cite{Perdomo-OrtizTURNmapping2010}.

The experiment consists of the following steps: a) construction of the energy function to be minimized in terms of the turn encoding; b) reduction of the energy expression to a two-body Hamiltonian; and finally, c) embedding in the device. These last two steps need additional resources as explained below. We will focus on the simplest example (Experiment 3, Fig.~\ref{fig:lanscape6AA}) to show the procedure in detail. The embeddings for the other five experiments are provided in the SI material.
The energy function for Experiment 3, containing the contributions due to on-site penalties for overlapping amino acids, and pairwise interactions between amino acids is,
\begin{equation}\label{eq:Eexp3}
E(\boldsymbol{q}) \equiv E^{cubic}_{exp3}=-1 - 4 q_{3} + 9 q_{1} q_{3} + 9 q_{2} q_{3} - 16 q_{1} q_{2} q_{3}
\end{equation}
where $q_1 0$ ($q_2 q_3$) encodes the orientation of the fourth (fifth) bond (see Fig.~\ref{fig:lanscape6AA}). From Eq.~\ref{eq:Eexp3} one can verify by substitution that the eight possible three-bit-variable assignments provide the desired energy landscape: the six conformations with $E\le 0$ shown in blue in Fig.~\ref{fig:lanscape6AA}.


Eq.~\ref{eq:Eexp3} describes the energy landscape of configurations but it is not quite ready for the device. Experimentally, we can specify up to two-body spin interactions, $\sigma^z_{i} \sigma^z_{j}$, and therefore, we need to convert this cubic energy function (Eq.~\ref{eq:Eexp3}) into a quadratic form resembling Eq.~\ref{eq:QUBO} (see SI for details). The resulting expression is
\begin{equation}
\begin{split}
H^{\text{unembedded}}_{p} &=(7 \sigma^z_1 + 9 \sigma^z_2 + 8 \sigma^z_3 - 20 \sigma^z_4+ 9 \sigma^z_1 \sigma^z_3 + 9 \sigma^z_2 \sigma^z_3  \\& -
 16 \sigma^z_1 \sigma^z_4 - 18 \sigma^z_2 \sigma^z_4 - 18 \sigma^z_3 \sigma^z_4)/4
\end{split}
\label{eq:h-isingFourVariable}
\end{equation}
where the original binary variables and spin operators are related by $q_i \rightarrow (1-\sigma^z_i)/2$. Experimental measurements of $\sigma^z_i$ yield $s_i = +1$ ($s_i= -1$) corresponding to $q_i  = 0$ ($q_i = 1$). Since $q_i = (1-s_i)/2$, measurement of $s_1$, $s_2$, and $s_3$ allows us to reconstruct the bit string $q_1 0 q_2 q_3$ which encodes the desired fold.

One ancilla variable was added during the transformation of the three-variable cubic Hamiltonian into this quadratic four-variable expression. The meaning of the original variables $s_1$, $s_2$, and $s_3$ remains the same, allowing for the reconstruction of the folds. The energy of this four-variable expression will not change as long as the measurements of $\sigma^z_1$ through $\sigma^z_4$ result in values for $q_1 q_2 q_3 q_4$ satisfying $q_4  = q_2 q_3$. This transformation ensures an energy penalty whenever this condition is violated.

\begin{figure}[H]
\begin{center}
\includegraphics[width=0.8\textwidth]{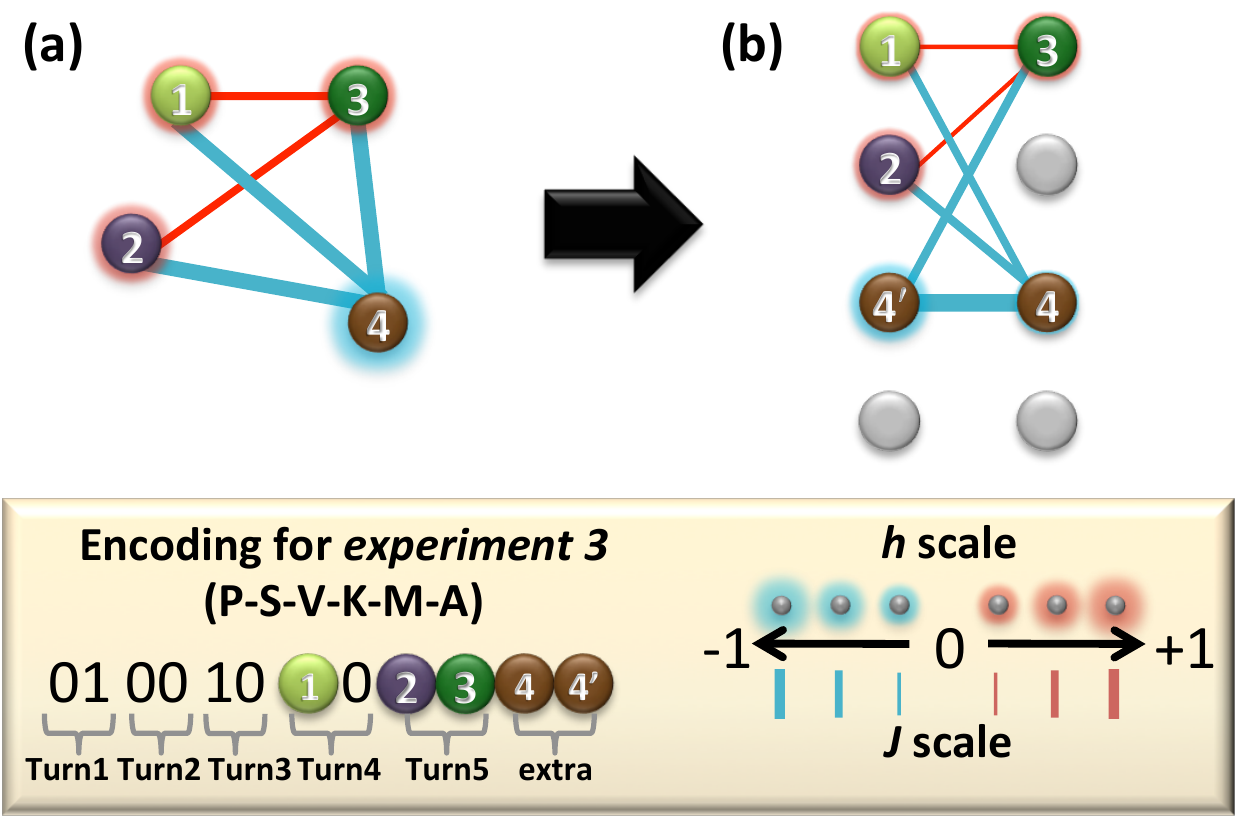}
\end{center}
\caption{Graph representations of (a) the four-qubit unembedded energy function (Eq.~\ref{eq:h-isingFourVariable}) and (b) the five-qubit expression (Eq.~\ref{eq:h-ising}) as was embedded into the quantum hardware. In graphs (a) and (b), each node denotes a qubit and the color and extent of its glow denotes the sign and strength of its corresponding longitudinal field, $h_i$. The edges represent the interaction couplings, $J_{ij}$, where color indicates sign and thickness indicates magnitude. Since we want the two qubits representing $q_4$ ($q_{4}$ and $q_{4^{\prime}}$) to end up with the same value, we apply the maximum ferromagnetic coupling ($J=-1$) between them, which adds a penalty whenever this equality is violated. These maximum couplings are indicated in the figure by heavy lines. For the case of Experiment 3, the reconstruction of the binary bit stings representing the folds in Fig.~\ref{fig:lanscape6AA}, from the five-quibt experimental measurements can be recovered by $\boldsymbol{q}_{exp3}= 0 1 0 0 1 0 q_1 0 q_{2} q_3 |q_{4}q_{4^{\prime}}$, with $q_i = 0$ ($q_i = 1$) whenever $s_i = 1$ ($s_i = -1$).}
\label{fig:ExpEmbed}
\end{figure}

The architecture of the chip lacks sufficient connectivity between the superconducting rings for a one-to-one assignment of variables to qubits (see Fig.~\ref{fig:ExpEmbed}). To satisfy the connectivity  requirements of the four-variable energy function, the couplings of one of the most connected variables, $q_4$, were fulfilled by duplicating this variable inside the device such that $q_4 \rightarrow q_{4}$ and $q_{4^{\prime}}$. In the form of Eq.~\ref{h-ising} the final expression representing the energy function of Experiment 3 is given by,
\begin{equation}
\begin{split}
H_{p} &=(7 \sigma^z_2\sigma^z_1+ 9 \sigma^z_2+ 8 \sigma^z_3+ 9 \sigma^z_1\sigma^z_3+ 9 \sigma^z_2\sigma^z_{4^{\prime}}- 2 \sigma^z_{4^{\prime}} -16 \sigma^z_1\sigma^z_{4^{\prime}} \\&
- 18 \sigma^z_2\sigma^z_{4^{\prime}}- 18 \sigma^z_{4^{\prime\prime}}- 18 \sigma^z_3\sigma^z_{4^{\prime\prime}}- 36 \sigma^z_{4^{\prime\prime}}\sigma^z_{4^{\prime\prime}})/36
\end{split}
\label{eq:h-ising}
\end{equation}
This expression satisfies all requirements for the problem Hamiltonian (Eq.~\ref{h-AQO}), the completion of which allows for the measurement of the energetic minimum conformation of this small peptide instance. The embedding of Eq.~\ref{eq:h-ising} into the hardware is shown in Fig.~\ref{fig:ExpEmbed}, where we label the five qubits used, $q_1$, $q_2$, $q_3$, $q_{4}$, and $q_{4^{\prime}}$. Since we want the two qubits representing $q_4$ to end up with the same value, we apply the maximum ferromagnetic coupling ($J=-1$) between them, which adds a penalty whenever this equality is violated  (last term in Eq.~\ref{eq:h-ising}). These maximum couplings are indicated in Fig.~\ref{fig:ExpEmbed} by heavy lines. The thinner lines show the remaining couplings used to realize the quadratic terms in Eq.~\ref{eq:h-ising}, color coded according to the sign of the interaction and its thickness representing their strength. Note that every quadratic term in Eq.~\ref{eq:h-ising} has a corresponding coupler. Hereafter, we will denote the outcome of the five-qubit measurements as $\boldsymbol{q}_{expo}= 010010q_10q_2q_3|q_{4} q_{4^{\prime}}$, with $q_i = 0$ ($q_i = 1$) whenever $s_i = 1$ ($s_i = -1$). Notice that only the bits preceding the divider character $|$ contain physical information. These are the ones shown under each of the protein fold drawings associated with Experiment 3 (see Fig.~\ref{fig:lanscape6AA}).


Similar embedding procedures to the one previously described were used for the larger experiments. For example, in Experiment 1, only 5 qubits define solutions of the computational problem. We needed 5 auxiliary qubits to transform the expression with 5-body interactions into an expression with only 2-body interactions. Embedding of this final expression required an additional of 18 qubits to satisfy the hardware connectivity requirements, for a total of 28 qubits. Table~I in the SI material summarizes the number of qubits required in each step through to the final experimental realizations.

Even though the quantum device follows a quantum annealing protocol, the odds of measuring the ground state are not necessarily high. For example, in the 81 qubit experiment, only 13 out of 10,000 measurements yielded the desired solution. We attribute these low-percentages to the 
analog nature of the device and to precision limitations in the real values of the local fields and couplings among the qubits in the experimental setup. When compared to other problem implementations, physical problems such as lattice folding lack the structure of the Ramsey number problem~\cite{Zhengbing}. In the lattice folding problem implemented here, the parameters defining the problem instances are arbitrary and do not fall into certain integral distinct values as in the case of the Ramsey number experiment, making precision issues more pronounced in our implementation.

To gain insights into the dynamics and evolution of the quantum system, we numerically simulated the superconducting array with a Bloch-Redfield model of the 8-qubit experiment (see SI material) which takes into account thermal fluctuations in the states due to the finite temperature (20mK) of the quantum device. For this 8-qubit experiment, the simulation predicted a ground state probability of 80.7 \%, in excellent agreement with the experimentally observed value (80.3\%). It is important to note that no adjustable parameters were used in our simulations to fit the data and all the parameters correspond to values measured directly from the quantum device. More details about the numerical simulations can be found in the SI. 

As seen in Fig.~\ref{fig:encoding}(c), the temperature of the device is comparable with the minimum gap of the eight-qubit Hamiltonian. Therefore, we expect stronger excitation/relaxation near the gap closing, $\tau \approx 0.6$, due to exchange of energy with the environment, when compared to the other regimes of the annealing schedule where the gap is much larger than $k_B T$. In the absence of environment (a fully coherent process), our simulations indicate that that the success probability would be 100\%, within numerical error. Fig.~\ref{fig:encoding}(d) shows that for the simulations at 20mK, the probability in the ground state goes down to $\sim 55\%$, but the same fluctuations make the system relax back to the ground state, yielding tan 80.27\% success probability. This is due to the advantageous natural tendency of the system to approach a thermal equilibrium which favors the ground state after crossing the minimum energy gap. As previously discussed in similar numerical simulations of quantum annealing algorithms~\cite{Amin_PRA2009a}, strong coupling to the bath and non-Markovianity would require going beyond the Bloch-Redfield model~\cite{Amin_PRA2009b}, but the agreement between experimental and simulated results support the validity of the quantum mechanical model used to describe the device. Previously reported temperature dependence predictions for the tunneling rate on the same qubits~\cite{johnson_quantum_2011} [3] and excellent agreement with the same level of theory used here reinforce the validity of our simulations for this 8-qubit instances.


We present the first quantum-mechanical implementation of lattice protein models using a programmable quantum device. We were able to encode and to solve the global minima solution for a small tetrapeptide and hexapeptide chain under several experimental schemes involving 5 and 8 qubits for the four-amino-acid sequence (Hydrophobic-Polar model) and 5, 27, 28, and 81 qubits experiments for the six amino-acid sequence under the Miyazawa-Jernigan model for general pairwise interactions. For the experiment with 8 qubits, we simulated the dynamics of the quantum device with a Redfield equation with no adjustable parameters, obtaining excellent agreement with experiment. Since the quantum annealing algorithm not only finds the ground state but also the low-lying excited states, it provides information about the relevant minimum energy compact structures of protein sequences~\cite{Camacho_PRL1993} and it is useful to evaluate designability and stability such as that found in natural protein sequences, where the global minimum of free energy is well separated in energy from other misfolded states~\cite{anfinsen_principles_1973}. The approach employed here can be extended to treat other problems in biophysics and statistical mechanics such as molecular recognition, protein design, and sequence alignment~\cite{hartmann2004}. 

\section*{Acknowledgements}

This work was supported by NSF CCI center, ``Quantum Information for Quantum Chemistry(QIQC)", Award number CHE-1037992. The authors thank Sergio Boixo, Mohammad Amin, and Ryan Babbush for helpful discussions and revisions of the manuscript.

\newpage

\newpage

\newpage

\newpage

\bibliographystyle{unsrt}

\pagebreak{}

\onecolumngrid

\section*{\setcounter{page}{1}}

\global\long\def\theequation{S\arabic{equation}}
 \setcounter{equation}{0}

\global\long\def\thefigure{S\arabic{figure}}
 \setcounter{figure}{0}

\begin{center}
\Large{\emph{Supplementary Information} }
\end{center}

\section*{Summary}
The paper \textit{Finding low-energy conformations of lattice protein models by quantum-annealing} presents the first experimental and largest quantum annealing experiment related to an optimization problem in the physical sciences. In Sec.~\ref{sec:energyfunction}, we summarize the construction of a more succinct version of the energy function describing the energy landscape of the six experimental realizations of the generalized lattice-folding model using Miyazawa-Jernigan pairwise interactions. In Sec.~\ref{sec:embedding}, we present the necessary steps to transform the energy function into an expression which can be readily implemented in the quantum device. In Sec.~\ref{sec:device}, we describe the quantum device used for our experiments and in Sec.~\ref{sec:qsimulations} we give details about the quantum simulations and results used to support the experimental outcomes.

\section{Transformation of the energy function of the lattice-folding model into the experimentally realizable spin-glass Hamiltonian}\label{sec:energyfunction}

The energy function for the lattice model can be obtained as a sum of different contributions,
\begin{equation}\label{eq:Etotal}
E_p(\boldsymbol{q}) = E_{onsite}(\boldsymbol{q}) + E_{pw}(\boldsymbol{q}) + E_{ext}(\boldsymbol{q})
\end{equation}
where $E_{onsite}(\boldsymbol{q})$ penalizes configurations with overlaps among any two amino acids, $E_{pw}(\boldsymbol{q})$ accounts for nearest-neighbor pairwise-interaction energies among non-bonded amino acids, and $E_{ext}(\boldsymbol{q})$ refers to any external potentials other than the ones coming from interactions among the amino acids defining the protein. For amino acid sequences \textit{in vacuo}, only $E_{onsite}$ and $E_{pw}$ are needed. The construction of these three-types of energy functions, in 2D and in 3D, for an arbitrary number of amino acids and interactions among them is explained in detail in Ref.~\onlinecite{Perdomo-OrtizTURNmapping2010}. Hereforth, we will only focus on the case of energy functions in 2D.

\subsection{Case of the six-amino acid sequence PSVKMA (Experiments 1-4)} 

\begin{figure}
\begin{center}
\includegraphics[width=0.95\textwidth]{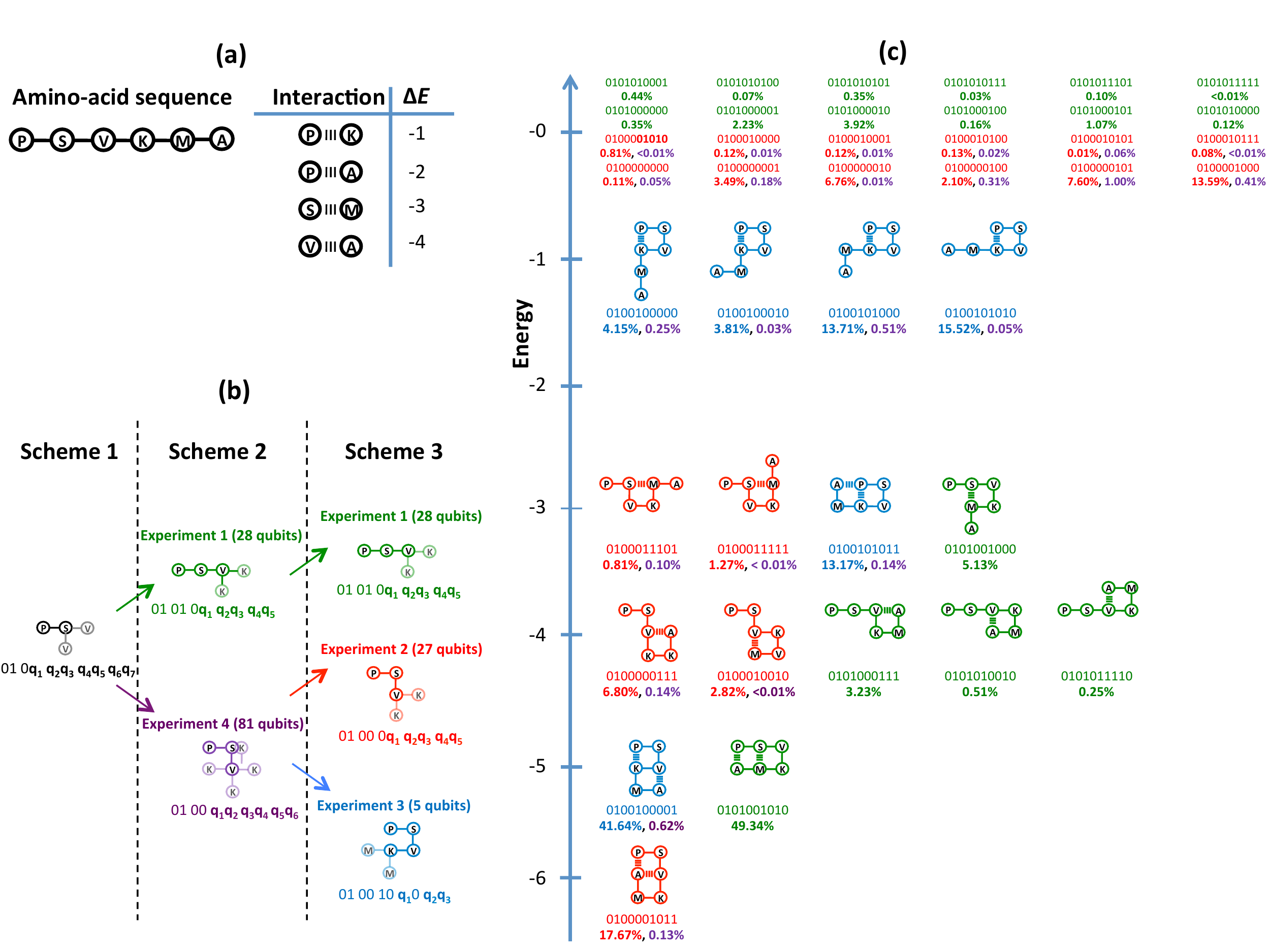}
\end{center}
\caption{(a) Representation of the six-amino acid sequence, Proline-Serine-Valine-Lysine-Methionine-Alanine with its respective one-letter sequence notation, PSVKMA. We use the pairwise nearest-neighbor Miyazawa-Jernigan interaction energies reported in Table 3 of Ref.~\onlinecite{miyazawa_residue-residue_1996}. (b) Divide and conquer approach showing three different schemes which independently solve the six-amino acid sequence PSVKMA on a two-dimensional lattice. We solved the problem under Scheme2 and 3 (Experiments 1 through 4). (c) Energy landscape for the valid conformations of the PSVKMA sequence. Results of the experimentally-measured probability outcomes are given as color-coded percentages according to each of the experimental realizations described in panel (b). Percentages for states with energy greater than zero are 32.70\%, 59.88\%, 8.00\%, and 95.97\% for Experiments 1 through 4, respectively.}
\label{fig:landscape6AA}
\end{figure}

For convenience, we reproduce Fig. 3 of the main text as Fig.~\ref{fig:landscape6AA}, which illustrates and defines the six amino-acid sequence PSVKMA.

As explained in the main text, the description of all possible 2D $N$-amino-acid fold in vacuo can be described by a bit string of length $2(N-1)$, with the first three bits held constant leaving $\ell = 2N -5$ binary variables as the computational variables of the problem,
\begin{equation}\label{eq:q}
\boldsymbol{q} = \underbrace{0 1}_{turn1} \underbrace{0 q_1}_{turn2} \underbrace{q_2 q_3}_{turn3} \cdots  \underbrace{q_{2N-6} q_{2N-5}}_{turn(N-1)}.
\end{equation}
For the case of $N=6$ (sequence PSVKMA), the problem is completely specified by the bit string
\begin{equation}\label{eq:q6AA}
\boldsymbol{q}_{6AA} = \underbrace{0 1}_{turn1} \underbrace{0 q_1}_{turn2} \underbrace{q_2 q_3}_{turn3} \underbrace{q_4 q_5}_{turn4} \underbrace{q_{6} q_{7}}_{turn5}.
\end{equation}

By using the construction in Ref.~\onlinecite{Perdomo-OrtizTURNmapping2010}, the 7-bit energy function describing the sequence PSVKMA (Scheme 1 in Fig.~\ref{fig:landscape6AA}) is given by,
\begin{equation}\label{eq:Epsvkma}
\begin{split}
E_{\textrm{\small{PSVKMA}}}&(\boldsymbol{q}_{6AA}) = -q_2 + 8 q_1 q_2 + 15 q_2 q_3 - 18 q_1 q_2 q_3 - 
3 q_1 q_4 + 12 q_1 q_2 q_4 + 4 q_3 q_4 + 3 q_1 q_3 q_4 \\& - 6 q_2 q_3 q_4
  - 12 q_1 q_2 q_3 q_4 + 4 q_2 q_5 +  3 q_1 q_2 q_5 - 15 q_2 q_3 q_5 + 15 q_4 q_5 
  +  3 q_1 q_4 q_5 \\& - 6 q_2 q_4 q_5 - 12 q_1 q_2 q_4 q_5  -  15 q_3 q_4 q_5 + 28 q_2 q_3 q_4 q_5  - 2 q_1 q_2 q_6 -  4 q_3 q_6 + 2 q_2 q_3 q_6 \\& + 13 q_1 q_2 q_3 q_6 -  2 q_1 q_4 q_6  + 4 q_1 q_2 q_4 q_6 + 2 q_3 q_4 q_6  +  13 q_1 q_3 q_4 q_6 + 4 q_2 q_3 q_4 q_6 \\&-  37 q_1 q_2 q_3 q_4 q_6  + 7 q_5 q_6 + 2 q_2 q_5 q_6 +  13 q_1 q_2 q_5 q_6 + 4 q_3 q_5 q_6 + 9 q_2 q_3 q_5 q_6 \\& -  33 q_1 q_2 q_3 q_5 q_6 - 20 q_4 q_5 q_6 +  13 q_1 q_4 q_5 q_6 + 4 q_2 q_4 q_5 q_6 -  37 q_1 q_2 q_4 q_5 q_6 + 9 q_3 q_4 q_5 q_6 \\& -  33 q_1 q_3 q_4 q_5 q_6 - 37 q_2 q_3 q_4 q_5 q_6   +  99 q_1 q_2 q_3 q_4 q_5 q_6 - 4 q_2 q_7 + 4 q_2 q_3 q_7 +  7 q_4 q_7 \\& + 2 q_2 q_4 q_7 + 13 q_1 q_2 q_4 q_7 +  4 q_3 q_4 q_7 + 9 q_2 q_3 q_4 q_7 -  33 q_1 q_2 q_3 q_4 q_7  + 4 q_2 q_5 q_7 \\& - 18 q_4 q_5 q_7 +  9 q_2 q_4 q_5 q_7 - 33 q_1 q_2 q_4 q_5 q_7 -  33 q_2 q_3 q_4 q_5 q_7 + 62 q_1 q_2 q_3 q_4 q_5 q_7 +  7 q_6 q_7 \\& + 2 q_2 q_6 q_7 + 13 q_1 q_2 q_6 q_7 +  4 q_3 q_6 q_7 + 9 q_2 q_3 q_6 q_7 -  33 q_1 q_2 q_3 q_6 q_7 - 20 q_4 q_6 q_7\\& +  13 q_1 q_4 q_6 q_7 + 4 q_2 q_4 q_6 q_7 -  37 q_1 q_2 q_4 q_6 q_7 + 9 q_3 q_4 q_6 q_7 -  33 q_1 q_3 q_4 q_6 q_7 - 37 q_2 q_3 q_4 q_6 q_7 \\& +  99 q_1 q_2 q_3 q_4 q_6 q_7 - 18 q_5 q_6 q_7 +  9 q_2 q_5 q_6 q_7 - 33 q_1 q_2 q_5 q_6 q_7 -  33 q_2 q_3 q_5 q_6 q_7  \\&  + 62 q_1 q_2 q_3 q_5 q_6 q_7 +  53 q_4 q_5 q_6 q_7 - 33 q_1 q_4 q_5 q_7 q_7 -  37 q_2 q_4 q_6 q_6 q_7+ 99 q_5 q_2 q_4 q_5 q_6 q_7 \\& -  33 q_3 q_1 q_5 q_6 q_7 + 62 q_1 q_4 q_4 q_5 q_6 q_7 +  99 q_2 q_3 q_4 q_5 q_6 q_7 -  190 q_1 q_2 q_3 q_4 q_5 q_6 q_7.
 \end{split}
\end{equation}

As shown in Fig.~\ref{fig:landscape6AA}, expressions for each of the different experiments in Schemes 2 and 3 can be sequentially obtained by fixing the value of some of the variables in $E_{\textrm{\small{PSVKMA}}}(\boldsymbol{q}_{6AA})$.

The energy function for Experiment 1 is obtained by evaluating $E_{\textrm{\small{PSVKMA}}}(\boldsymbol{q}_{6AA})$ with $q_1 =1$ (third amino-acid moves to the right) and $q_2 = 0$ (fourth amino-acid moves either down or right, exploiting upper/lower half-plane symmetry). After relabeling the five remaining variables so that their labels go from 1-5 instead of 3-7, i.e., $\boldsymbol{q}_{6AA} = 0 1 0 q_1 q_2 q_3  q_4 q_5 q_6 q_7 \xrightarrow[\rm{relabel}]{q_1=1,q_2=0} \boldsymbol{q}_{exp1} = 0 1 0  1 0 q_1 q_2 q_3  q_4 q_5$, the resulting expression describing the energy landscape for Experiment 1 is given by 
\begin{equation}\label{eq:Eexp1_5local}
\begin{split}
E^{exp1}_{\textrm{\small{PSVKMA}}}&(\boldsymbol{q}_{exp1}) = -3 q_2 + 7 q_1 q_2 + 18 q_2 q_3 - 15 q_1 q_2 q_3  -  4 q_1 q_4 - 2 q_2 q_4 + 15 q_1 q_2 q_4 \\&+ 7 q_3 q_4 +  4 q_1 q_3 q_4 - 7 q_2 q_3 q_4 - 24 q_1 q_2 q_3 q_4  +  7 q_2 q_5 + 4 q_1 q_2 q_5 - 18 q_2 q_3 q_5 \\& + 7 q_4 q_5 + 
 4 q_1 q_4 q_5 - 7 q_2 q_4 q_5 - 24 q_1 q_2 q_4 q_5 -  18 q_3 q_4 q_5 + 20 q_2 q_3 q_4 q_5 \\&+  29 q_1 q_2 q_3 q_4 q_5
 \end{split}
\end{equation}

The energy function for Experiment 4 is obtained by evaluating $E_{\textrm{\small{PSVKMA}}}(\boldsymbol{q}_{6AA})$ with $q_1 =0$ (third amino-acid moves down). After renaming the six remaining variables so that their labels span 1-6 instead of 2-7, i.e., $\boldsymbol{q}_{6AA} = 0 1 0 q_1 q_2 q_3  q_4 q_5 q_6 q_7 \xrightarrow[\rm{relabel}]{q_1=0} \boldsymbol{q}_{exp4} = 0 1 0  0 q_1 q_2 q_3  q_4 q_5 q_6$, the resulting expression describing the energy landscape for Experiment 4 is given by 
\begin{equation}\label{eq:Eexp4_6local}
\begin{split}
E^{exp4}_{\textrm{\small{PSVKMA}}}&(\boldsymbol{q}_{exp4}) = -q_1 + 15 q_1 q_2 + 4 q_2 q_3 - 6 q_1 q_2 q_3 + 4 q_1 q_4 -  15 q_1 q_2 q_4 + 15 q_3 q_4 - 6 q_1 q_3 q_4 \\& -  15 q_2 q_3 q_4 
+ 28 q_1 q_2 q_3 q_4 - 4 q_2 q_5 +  2 q_1 q_2 q_5 + 2 q_2 q_3 q_5 + 4 q_1 q_2 q_3 q_5 +  7 q_4 q_5 \\& + 7 q_5 q_6  + 2 q_1 q_4 q_5 + 4 q_2 q_4 q_5 +  9 q_1 q_2 q_4 q_5 - 20 q_3 q_4 q_5 + 4 q_1 q_3 q_4 q_5 +  9 q_2 q_3 q_4 q_5 \\& - 37 q_1 q_2 q_3 q_4 q_5 - 4 q_1 q_6 +  4 q_1 q_2 q_6 + 7 q_3 q_6 + 2 q_1 q_3 q_6 +  4 q_2 q_3 q_6 + 9 q_1 q_2 q_3 q_6 \\&+ 4 q_1 q_4 q_6 -  18 q_3 q_4 q_6 + 9 q_1 q_3 q_4 q_6 -  33 q_1 q_2 q_3 q_4 q_6 + 2 q_1 q_5 q_6 +  4 q_2 q_5 q_6 - 20 q_3 q_5 q_6  \\& + 9 q_1 q_2 q_5 q_6 +  4 q_1 q_3 q_5 q_6 + 9 q_2 q_3 q_5 q_6 -  37 q_1 q_2 q_3 q_5 q_6 - 18 q_4 q_5 q_6 +  9 q_1 q_4 q_5 q_6 \\& - 33 q_1 q_2 q_4 q_5 q_6 +  53 q_3 q_4 q_5 q_6 - 37 q_1 q_3 q_4 q_5 q_6 -  33 q_2 q_3 q_4 q_5 q_6 + 99 q_1 q_2 q_3 q_4 q_5 q_6
 \end{split}
\end{equation}

The energy function for Experiment 2 is obtained by evaluating $E^{exp4}_{\textrm{\small{PSVKMA}}}(\boldsymbol{q}_{exp4})$ with $q_1 =0$ (fourth amino-acid moves either down or right). After renaming the five remaining variables so that their labels span 1-5 instead of 2-6, i.e., $\boldsymbol{q}_{exp4} = 0 1 0  0 q_1 q_2 q_3  q_4 q_5 q_6 \xrightarrow[\rm{relabel}]{q_1=0} \boldsymbol{q}_{exp2} = 0 1 0  0 0 q_1 q_2 q_3  q_4 q_5$, the resulting expression describing the energy landscape for Experiment 2 is given by 
\begin{equation}\label{eq:Eexp2_5local}
\begin{split}
E^{exp2}_{\textrm{\small{PSVKMA}}}&(\boldsymbol{q}_{exp2}) = 4 q_1 q_2 + 15 q_2 q_3 - 15 q_1 q_2 q_3 - 4 q_1 q_4 +  2 q_1 q_2 q_4 + 7 q_3 q_4 + 4 q_1 q_3 q_4  \\&-  20 q_2 q_3 q_4 + 9 q_1 q_2 q_3 q_4 + 7 q_2 q_5 +  4 q_1 q_2 q_5 - 18 q_2 q_3 q_5 + 7 q_4 q_5 +  4 q_1 q_4 q_5 \\& - 20 q_2 q_4 q_5 + 9 q_1 q_2 q_4 q_5 -  18 q_3 q_4 q_5 + 53 q_2 q_3 q_4 q_5 -  33 q_1 q_2 q_3 q_4 q_5.
 \end{split}
\end{equation}
 
Finally, the energy function for Experiment 3 is obtained by evaluating $E^{exp4}_{\textrm{\small{PSVKMA}}}(\boldsymbol{q}_{exp4})$ with $q_1 =1$, $q_2 =0$(fourth amino-acid moves left) and $q_4 = 0$ (fifth amino-acid moves either down or left), exploiting the constrains imposed by the three fixed amino-acids (P,S, and V). After renaming the three remaining variables so that their labels are $q_1, q_2$ and $q_3$ instead of $q_3, q_5$, and $q_7$, i.e., $\boldsymbol{q}_{exp4} = 0 1 0 0 q_1 q_2 q_3  q_4 q_5 q_6 \xrightarrow[\rm{relabel}]{q_1=1,q_2=0, q_4=0} \boldsymbol{q}_{exp3} = 0 1 0   0 1 0 q_1 0 q_2  q_3$, the resulting expression describing the energy landscape for Experiment 3 is given by 
\begin{equation}\label{eq:Eexp3_3local}
E^{exp3}_{\textrm{\small{PSVKMA}}}(\boldsymbol{q}_{exp3}) = -1 - 4 q_3 + 9 q_1 q_3 + 9 q_2 q_3 - 16 q_1 q_2 q_3 
\end{equation}

\subsection{Case of the four-amino acid sequence HPPH (Experiment 5)} 

Besides the six-amino acid sequence considered above, we also constructed the energy function for the simplest of all sequences within lattice protein models, the HPPH four-amino acid sequence within the HP model. For $N=4$, we can specify any of its folds by the bit string $\boldsymbol{q}_{exp5} = 0 1 0 q_1 q_2 q_3$. The three-bit energy function describing the energy landscape of Experiment 5 (see Fig.~\ref{fig:landscapeHPPH}) is given by,
\begin{equation}\label{eq:Ehpphvacuo_3local}
E_{HPPH}(\boldsymbol{q}_{exp5}) = -q_2 + 2 q_1 q_2 + 2 q_2 q_3 - 3 q_1 q_2 q_3
\end{equation}
  
\subsection{Case of the four-amino acid sequence HPPH under external constraints (Experiment 6)} 

\begin{figure}
\begin{center}
\includegraphics[width=0.95\textwidth]{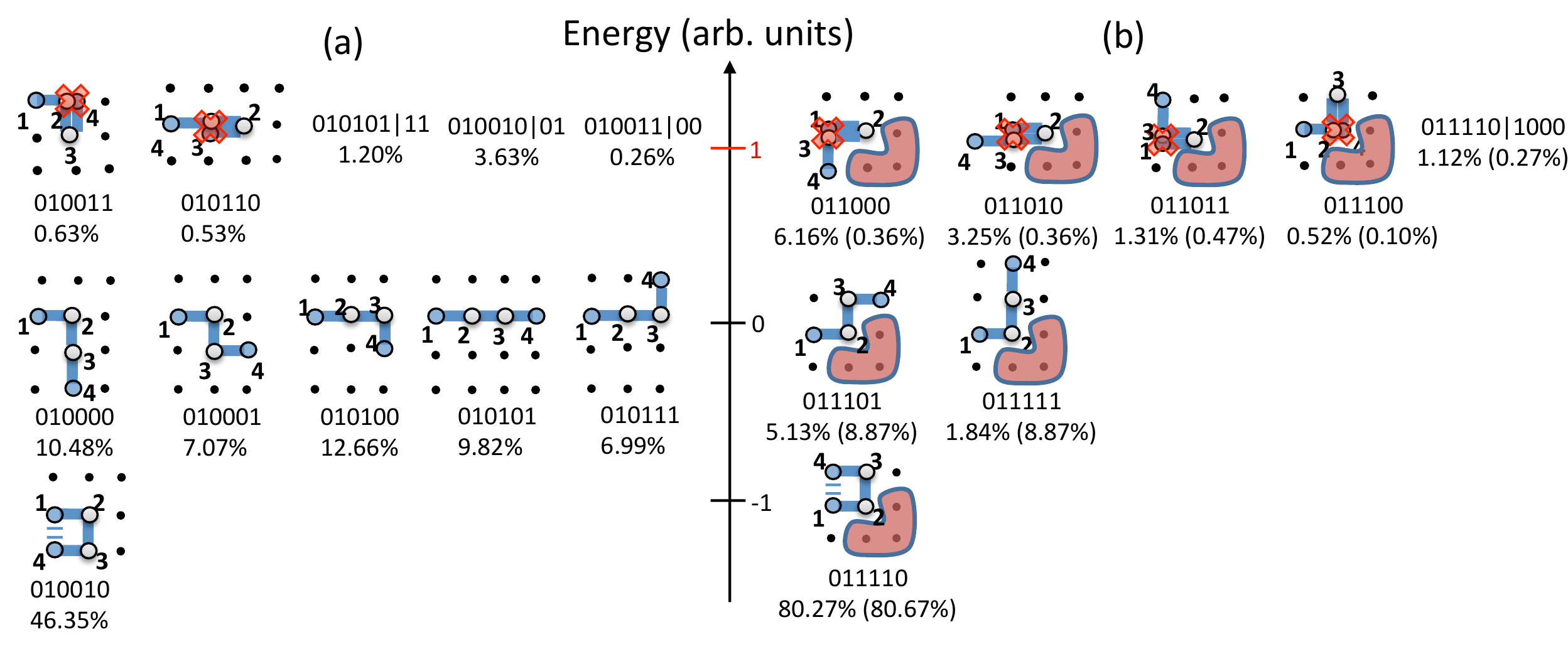}
\end{center}
\caption{\label{fig:landscapeHPPH}  
Energy landscape for the four amino-acid sequence HPPH, (a) \textit{in vacuo} (Experiment 5),  and (b) under the presence of a chaperone-like environment (Experiment 6) represented by the red-shaded region. In panel (a) [panel (b)], percentages indicate the experimentally measured probabilities of each state collected over 10,000 [28,672] runs of the quantum annealing algorithm described in Sec.~\ref{section:exp}. In the case of Experiment 6, numerical results from the Bloch-Redfield model discussed in Sec~\ref{sec:qsimulations} are included in parenthesis. Although the variables involved in Experiment 5 (Experiment 6) are described by $\boldsymbol{q}_{exp5}= 0 1 0 q_1 q_{2} q_3 | q_{4} q_{4^\prime}$ ($\boldsymbol{q}_{exp6}= 0 1 q_1 q_{2} q_3 q_{4} |q_5  q_6  q_{2^{\prime}} q_{4^{\prime}}$), under each fold we write only the physically-relevant variables which define the conformation. Since we show some experimental outcomes also for states with $E>0$, then it is natural to find states which violate either the \textsc{and} condition or the ferromagnetic condition; for these cases we explicitly write the auxilliary variables which went into the quantum hardware. For example, in Experiment 5 [panel (a)], the state $\boldsymbol{q}_{exp5} =  010101|11$ violates the \textsc{and} condition since $q_4 \neq q_2 q_3$. In the case of the state $\boldsymbol{q}_{exp5} =  010010|01$ the ferromagnetic condition for $q_4$ is violated since $q_4 \neq q_{4^\prime}$. Each overlap of the amino acids with the chaperone raises energy by four units, whereas overlaps (red crossings) among amino acids in the chain raise energy by two units.} 
\end{figure}

A more realistic \textit{in vivo} picture involves the presence of ÒchaperoneÓ proteins assisting the folding dynamics towards the global minima. Chaperones, molecular docking, and molecular recognition are examples of problems which can be studied by adding external potentials, $E_{ext}(\boldsymbol{q})$, beyond the intrinsic interactions defined by the amino-acid chain, $E_{onsite}(\boldsymbol{q})$ and $E_{pw}(\boldsymbol{q})$ (see Eq.~\ref{eq:Etotal}). The first consequence of adding an external potential $E_{ext}(\boldsymbol{q})$ (as the Òchaperone-likeÓ environment surrounding the small four-amino-acid sequence HPPH, illustrated in Fig.~\ref{fig:landscapeHPPH} by the pink-shaded area near the peptide)  is that we can no longer exploit the symmetry of the solution space for upper and lower half plane conformations. Therefore, we cannot set the first variable of the turn associated with the third amino-acid to zero. Under external potentials, we specify arbitrary folds of the four-amino acid problem by $\boldsymbol{q}_{exp6} = 0 1 q_1 q_2 q_3 q_4$, where $q_1 q_2 (q_3 q_4)$ encodes the orientation of the second (third) bond. 

The external potential penalizes conformations in which either the third or fourth amino acid go into the chaperone region is:
\begin{equation}\label{eq:Eext}
\begin{split}
E_{chap}(\boldsymbol{q}_{exp6}) &=\lambda^{d}_{ext} (1 - q_1) (1 - q_2) + \lambda^{r}_{ext} (1 - q_1) q_2 + \lambda^{dr}_{ext} (1 - q_1) (1 - q_2) (1 - q_3) q_4  \\& + \lambda^{rd}_{ext} (1 - q_1) q_2 (1 - q_3) (1 - q_4) 
\end{split}
\end{equation}
The penalty $\lambda^{d}_{ext}$ raises energy only when the third amino-acid moves down ($q_1=0, q_2=0$), $\lambda^{r}_{ext}$ raises energy only when the  third amino-acid moves right ($q_1=0, q_2=1$), $\lambda^{dr}_{ext}$ raises energy only when the third amino-acid moves down \emph{and} the fourth-amino acid moves right ($q_1=0, q_2=0,q_3=0, q_4=1$), and the last penalty, $\lambda^{rd}_{ext}$, raises energy only when the third amino-acid moves down \emph{and} the fourth-amino acid moves right ($q_1=0, q_2=1,q_3=0, q_4=0$). Each overlap of the amino acids with the chaperone increases energy by four units, i.e., $\lambda^{d}_{ext} =\lambda^{r}_{ext} = \lambda^{dr}_{ext} =\lambda^{rd}_{ext} = 4$.

When the third amino acid is also allowed to move upwards, the energy function for the HPPH chain in vacuo is given by,
\begin{equation}\label{eq:Ehpphvacuo_4local}
\begin{split}
E_{HPPH}(\boldsymbol{q}_{exp6}) &= q_1 - q_3 + q_1 q_3 + 2 q_2 q_3 - 4 q_1 q_2 q_3 + 
 2 q_1 q_4  \\& - 3 q_1 q_2 q_4 + 2 q_3 q_4 - 4 q_1 q_3 q_4 - 
 3 q_2 q_3 q_4 + 7 q_1 q_2 q_3 q_4
\end{split}
\end{equation} 
After adding Eq.~\ref{eq:Eext} and Eq.~\ref{eq:Ehpphvacuo_4local}, the resulting energy function for the HPPH peptide in the presence of the ``chaperone" environment illustrated in Fig.~\ref{fig:landscapeHPPH}, is given by,
\begin{equation}\label{eq:Ehpphenv_4local}
\begin{split}
E_{HPPH,chap}(\boldsymbol{q}_{exp6}) &= 4 - 3 q_{1} + 4 q_{2} - 4 q_{1} q_{2} - q_{3} + q_{1} q_{3} - 2 q_{2} q_{3} +  4 q_{4} - 2 q_{1} q_{4} \\&  - 8 q_{2} q_{4} + 5 q_{1} q_{2} q_{4} - 2 q_{3} q_{4} + 5 q_{2} q_{3} q_{4} - q_{1} q_{2} q_{3} q_{4},
\end{split}
\end{equation}

\section{Embbedding of problem instances into the quantum hardware}\label{sec:embedding}

\subsection{Reduction of high-order terms to a 2-body Ising-like Hamiltonian}
As explained in the main text, although the above energy expressions (Eqs. \ref{eq:Epsvkma}S, \ref{eq:Eexp1_5local}S, \ref{eq:Eexp4_6local}S, \ref{eq:Eexp2_5local}S, \ref{eq:Eexp3_3local}S, \ref{eq:Ehpphvacuo_3local}S, and \ref{eq:Ehpphenv_4local}S) describe the desired energy landscape, they are not suitable for experimental implementation. We need to reduce the degree of the high-order terms (cubic, cuartic, etc) to a quadratic expression (up to 2-body interactions). These high-order terms indicate many-body interactions which are not experimentally feasible within the current quantum device. To achieve this without altering the low-energy spectra ($E \le 0$) where the target minima is supposed to be found, we use the technique described in Ref.~\onlinecite{Perdomo2008,Perdomo-OrtizTURNmapping2010}. In the main text, we presented the simplest case where only one reduction was required (expression for Experiment 3, Eq.~\ref{eq:Eexp3_3local}). In the following we will focus on the next most complex case (Experiment 6, Eq.~\ref{eq:Ehpphenv_4local}) which can be easily generalized to obtain any of the 2-body energy expressions for the larger experiments.

We introduce two ancilla binary variables, $q_5$ and $q_6$, and substitute Eq.~\ref{eq:Ehpphenv_4local} with products of the form $q_1 q_2$ into $q_5$ and $q_3 q_4$ into $q_6$. This substitution transforms the energy expression (Eq.~\ref{eq:Ehpphenv_4local}) into a quadratic expression, e.g, the highest-order term which is quadratic, $q_1 q_2 q_3 q_4$, is replaced by $q_5 q_6$ which becomes quadratic, as desired. Under these substitutions, whenever we have six-variable assignments, $\boldsymbol{q}_{exp6} = 01 q_1 q_2 q_3 q_4 | q_5 q_6$,  such that $q_6 = q_1 \wedge q_2 = q_1 q_2$ and $q_6 = q_3 \wedge q_4 = q_3  q_4$, we have the same energy spectrum as the one for the original quartic, four-variable expression. Since these two ancilla are new variables whose values are independent of the four original variables, we need to penalize six-variable assignments whenever $q_5 \neq q_1 q_2$ and $q_6 \neq q_3 q_4$. For every ``collapse" of the form $q_i q_j \rightarrow r_k$, we add the penalty $E_{\wedge}(q_i, q_j, r_k;\delta_{ij}) = \delta_{ij} (3 r_k+ q_i q_j - 2 q_i r_k - 2 q_j \delta_{ij})$, where $\delta_{ij}$ is a positive number representing a penalty chosen (for more details see Ref.~\onlinecite{Perdomo-OrtizTURNmapping2010}) such that assignments violating this \textsc{and} condition correspond to free-energies $E>0$, outside the relevant search region ($E \le 0$). The function $E_{\wedge}(q_i, q_j, r_k;\delta_{ij}) = 0$ only if $r_k = q_i q_j$ \emph{and} $E_{\wedge}(q_i, q_j, r_k;\delta_{ij}) > 0$ if $r_k \neq q_i q_j$. The six-variable expression resulting from the insertion of the new ancilla variables plus the penalty function becomes,
\begin{equation}
\begin{split}
E_{HPPH,chap}^{2body}(\boldsymbol{q}) &= E_{HPPH,chap}(q_1, q_2, q_3, q_4; q_1q_2 \rightarrow q_5, q_3q_4 \rightarrow q_6) \\&+ E_{\wedge}(q_1, q_2, q_5;\delta_{12}) + E_{\wedge}(q_3, q_4, q_6;\delta_{34}) \\ &= 4 - 3 q_{1} + 4 q_{2} + 6 q_{1} q_{2} - q_{3} + q_{1} q_{3} - 2 q_{2} q_{3} +  4 q_{4} - 2 q_{1} q_{4} - 8 q_{2} q_{4} + 4 q_{3} q_{4} \\& + 14 q_{5} - 12 q_{1} q_{5} - 12 q_{2} q_{5} + 5 q_{4} q_{5} + 10 q_{6} + 5 q_{2} q_{6} - 8 q_{3} q_{6} - 8 q_{4} q_{6} - q_{5} q_{6}
\end{split}
\end{equation}
where according to the criteria in Ref.~\onlinecite{Perdomo-OrtizTURNmapping2010}, we have chosen $\delta_{12} = 6$, and $\delta_{34} = 4$.

To rewrite this quadratic form in terms of the spin variables $\{s_i\}$, we apply the transformation $q_i \equiv \frac{1}{2} (1 - s_i)$ to each of the binary variables,
\begin{equation}
\begin{split}
E_{HPPH,chap}^{Ising} &= 10 + \frac{13}{4} s_{1} + \frac{3}{4} s_{2}  + \frac{7}{4} s_{3} + \frac{1}{4} s_{4} - 2 s_{5} - 2 s_{6} + \frac{3}{2} s_{1} s_{2} +
 \frac{1}{4} s_{1} s_{3} - \frac{1}{2} s_{2} s_{3}
  \\&   - \frac{1}{2} s_{1} s_{4} -
 2 s_{2} s_{4} + s_{3} s_{4}  - 3 s_{1} s_{5} - 3 s_{2} s_{5} +
 \frac{5}{4} s_{4} s_{5}  + \frac{5}{4} s_{2} s_{6} - 2 s_{3} s_{6} \\& - 2 s_{4} s_{6} -
 \frac{1}{4} s_{5} s_{6}
\end{split}
\label{eq:e-ising}
\end{equation}
After substracting the constant (independent term), we can fulfill the requirement that $\abs{h_i}\le1$ and $\abs{J_{ij}}\le1$ by scaling all coefficients of Eq.~\ref{eq:e-ising} down by the maximum absolute value of all coefficients. The renormalized quadratic expression is given by,
\begin{equation}
\begin{split}
E_{exp6}^{unembedded}(\boldsymbol{s}) &= \frac{4}{13}(E_{HPPH,chap}^{ising}-10) \\&= (13 s_{1} + 3 s_{2} + 7 s_{3} + s_{4}  - 8 s_{5} - 8 s_{6} + 6 s_{1} s_{2}  +
  s_{1} s_{3} - 2 s_{2} s_{3}
  \\&   - 2 s_{1} s_{4} -
 8 s_{2} s_{4} +  4 s_{3} s_{4}  - 12 s_{1} s_{5} - 12 s_{2} s_{5} +
 5 s_{4} s_{5}  + 5 s_{2} s_{6} - 8 s_{3} s_{6} \\& - 8 s_{4} s_{6} -
  s_{5} s_{6})/13
\end{split}
\end{equation}
The final Ising spin-glass Hamiltonian (before embedding into the quantum device) can be obtained by the substitution $s_i \rightarrow \sigma^z_i$.
\begin{equation}\label{eq:Hhpphenv_unembed}
\begin{split}
H^{\text{unembedded}}_{exp6} &= (13 \sigma^z_1 + 3 \sigma^z_{2} + 7 \sigma^z_3 + \sigma^z_{4} - 8 \sigma^z_5 - 8 \sigma^z_6 + 6 \sigma^z_1 \sigma^z_{2}  + \sigma^z_1 \sigma^z_3 - 2 \sigma^z_{2} \sigma^z_3
  \\& - 2 \sigma^z_1 \sigma^z_{4} -
 8 \sigma^z_{2} \sigma^z_{4} + 4 \sigma^z_3 \sigma^z_{4} - 12 \sigma^z_1 \sigma^z_5 - 12 \sigma^z_{2} \sigma^z_5 +
 5 \sigma^z_{4} \sigma^z_5  + 5 \sigma^z_{2} \sigma^z_6 - 8 \sigma^z_3 \sigma^z_6 \\& - 8 \sigma^z_{4} \sigma^z_6 -
  \sigma^z_5 \sigma^z_6)/13
\end{split}
\end{equation}

\subsection{Embedding into the quantum hardware}
\begin{figure}[h]
\begin{center}
\includegraphics[width=0.9\textwidth]{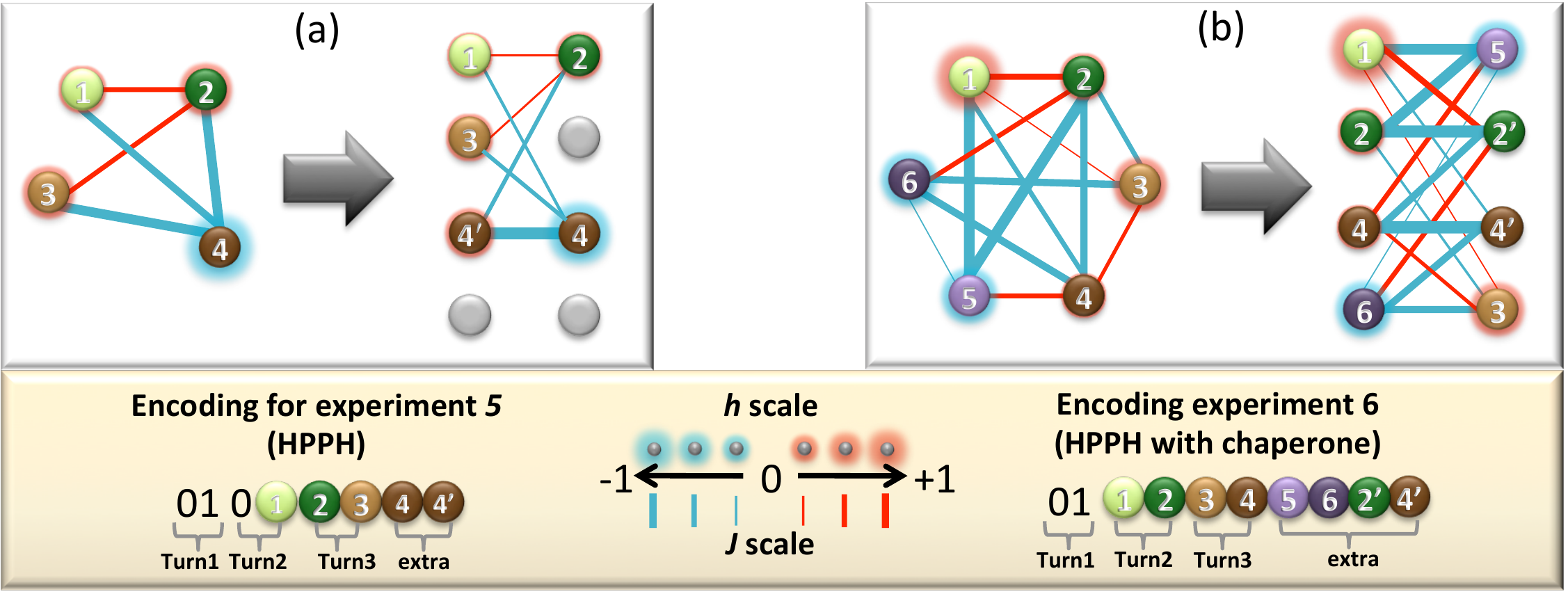}
\end{center}
\vspace{-0.5cm}
\caption{\label{fig:embeddingExp5Exp6} In the graphs presented in (a) and (b), each node denotes a qubit, and the color and extent of its glow denotes the sign and strength of its corresponding longitudinal field, $h_i$. The edges represent the interaction couplings, $J_{ij}$, where color indicates sign and thickness indicates magnitude. The maximum couplings are indicated in the figure by heavy lines. (a) Primal graph (left) and the embedded representation of the expression implemented in the quantum hardware for Experiment 5 (HPPH \textit{in vacuo}). (b) Primal graph (left) and the embedded eight-qubit expression (Eq.~\ref{eq:h_embeddingExp6}) for Experiment 6 (HPPH in the chaperone-lke environment).}
\end{figure}

Eq.~\ref{eq:Hhpphenv_unembed} does not fulfill the chip-connectivity requirements (see Fig.~\ref{fig:embeddingExp5Exp6}) for the primal graph representing Eq.~\ref{eq:Hhpphenv_unembed}. This limitation is fixed at the cost of adding two new qubits serving as replicas of the two qubits which are linked by more than four connections. To enforce that the replicas of the $i$-th qubit ($\sigma^z_{i^{\prime}}$) produce the same outcome as the original $i$-th qubit, we couple $\sigma^z_{i}$ and $\sigma^z_{i^{\prime}}$ with a strong ferromagnetic coupling, such that whenever the outcomes of the two variables are different they get penalized by a chosen penalty factor $\gamma_i > 0$. The function which performs this penalization for each replica $i$-th qubit is $H_{FM}(\{\sigma^z_{i}\};\gamma_i) = \gamma_i (1-\sigma^z_{i}
\sigma^z_{i^{\prime}})$. Notice that $H_{FM}(\{\sigma^z_{i}\};\gamma_i) = 0$, if $s_{i}  = s_{i^{\prime}}  = \pm 1$, but $H_{FM}(\{\sigma^z_{i}\};\gamma_i) = 2 \gamma_i$, if $s_{i} \neq s_{i^{\prime}}$. For this study, a value of $\gamma_2 = \gamma_4 = 1$ suffices to leave assignments which violate this condition outside the region of interest with $E \le 0$.

The redistribution of the connections among the original and primed qubits is given in the right panel of Fig.~\ref{fig:embeddingExp5Exp6}. The modified function taking into account the added ferromagnetic couplings is,
\begin{equation}
\begin{split}
\tilde{H}_{exp6} &= H^{\text{unembedded}}_{exp6}(\sigma^z_2 \rightarrow \{\sigma^z_{2},\sigma^z_{2^{\prime}} \};\sigma^z_4 \rightarrow \{\sigma^z_{4},\sigma^z_{4^{\prime}}\}) \\& + H_{FM}(\{\sigma^z_{2}\};\gamma_2=1) + H_{FM}(\{\sigma^z_{4}\};\gamma_4 =1) \\& = (13 \sigma^z_1 + 3 \sigma^z_{2} + 7 \sigma^z_3 + \sigma^z_{4} - 8 \sigma^z_5 - 8 \sigma^z_6 + 6 \sigma^z_1 \sigma^z_{2^\prime}  + \sigma^z_1 \sigma^z_3 - 2 \sigma^z_{2} \sigma^z_3
  \\& - 2 \sigma^z_1 \sigma^z_{4^{\prime}} -
 8 \sigma^z_{2^\prime} \sigma^z_{4} + 4 \sigma^z_3 \sigma^z_{4} - 12 \sigma^z_1 \sigma^z_5 - 12 \sigma^z_{2} \sigma^z_5 +
 5 \sigma^z_{4} \sigma^z_5  + 5 \sigma^z_{2^{\prime}} \sigma^z_6 - 8 \sigma^z_3 \sigma^z_6 \\& - 8 \sigma^z_{4^{\prime}} \sigma^z_6 -
  \sigma^z_5 \sigma^z_6)/13  + (1-\sigma^z_{2}
\sigma^z_{2^{\prime}}) + (1-\sigma^z_{4}
\sigma^z_{4^{\prime}})
\end{split}
\label{eq:hTilde_embeddingExp6}
\end{equation}
Again, we subtract the independent constant terms from the insertion of the $H_{FM}$ functions. The final expression, which is implementable in the quantum device is,
\begin{equation}
\begin{split}
H_{exp6} &= \tilde{H}_{exp6} - 2 = (13 \sigma^z_1 + 3 \sigma^z_{2} + 7 \sigma^z_3 + \sigma^z_{4} - 8 \sigma^z_5 - 8 \sigma^z_6 + 6 \sigma^z_1 \sigma^z_{2^\prime}  + \sigma^z_1 \sigma^z_3 - 2 \sigma^z_{2} \sigma^z_3
  \\& - 2 \sigma^z_1 \sigma^z_{4^{\prime}} -
 8 \sigma^z_{2^\prime} \sigma^z_{4} + 4 \sigma^z_3 \sigma^z_{4} - 12 \sigma^z_1 \sigma^z_5 - 12 \sigma^z_{2} \sigma^z_5 +
 5 \sigma^z_{4} \sigma^z_5  + 5 \sigma^z_{2^{\prime}} \sigma^z_6 - 8 \sigma^z_3 \sigma^z_6 \\& - 8 \sigma^z_{4^{\prime}} \sigma^z_6 -
  \sigma^z_5 \sigma^z_6 - 13 \sigma^z_{2} \sigma^z_{2^{\prime}} - 13 \sigma^z_{4} \sigma^z_{4^{\prime}})/13
\end{split}
\label{eq:h_embeddingExp6}
\end{equation}

\begin{table}[H]
 \caption{\label{table:numberqbits} Number of qubits needed for each one of the six experiments described in Fig.~\ref{fig:landscape6AA} and Fig.~\ref{fig:landscapeHPPH}. The most compact version of the energy function corresponds to the second column. Each one of the steps, reduction of high-order terms in the energy function towards a 2-body Ising-like Hamiltonian and embedding of this Ising expression to fulfill the physical connectivity of the qubits in the device, requires more auxiliary qubits. The final column reports the number of qubits in the experimentally implemented expression of the energy function.}
\begin{ruledtabular}
\begin{tabular}{c|c|c|c}
   & \multicolumn{3}{c}{\bfseries Number of qubits needed} \\
 \hline
 Experiment \# & energy function & Ising Hamiltonian & hardware-embedded expression \\ 
 \hline\hline
 1 & 5 & 10 &  28\\
 2 & 5 & 10 &  27\\
 3 & 3 & 4 &  5\\
 4 & 6 & 19 &  81\\
 5 & 3 & 4 &  5\\
 6 & 4 & 6 &  8\\
 \end{tabular}
\end{ruledtabular}
\end{table}

\begin{figure}[h]
\begin{center}
\includegraphics[width=0.8\textwidth]{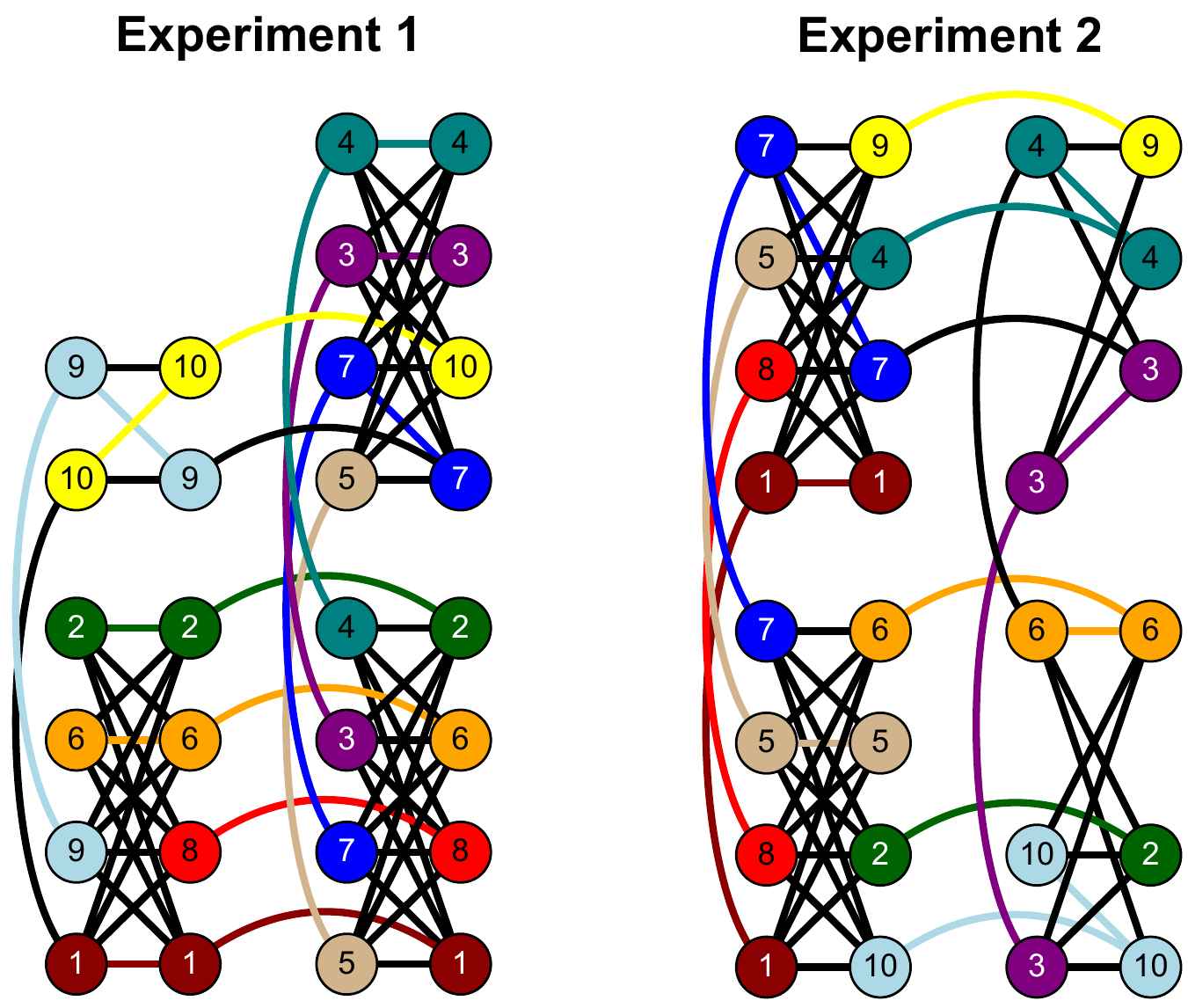}
\end{center}
\caption{Embedding of Experiments 1 and 2 into the quantum hardware. The 28 qubits (27 qubits) from Experiment 1 (Experiment 2) have been relabeled to show the qubits which were strongly ferromagnetically coupled representing the same variable and biased to have the same experimental outcomes. Both problem instances resulted in ten-qubit spin-glass Hamiltonians after reducing their energy expressions to the Ising-like 2-body interaction expression. The additional qubits are part of the embedding procedure used to fulfill the arbitrary connectivity of the Ising expression, allowing for couplings of qubits that are not directly coupled in hardware. }
\label{fig:embeddingsExp1Exp2}
\end{figure}

The embeddings for Experiments 3 and 4 are shown in Fig.~4 and Fig.~1 of the main text, respectively. The embeddings corresponding to Experiment 5 and 6 are represented in Fig.~\ref{fig:embeddingExp5Exp6}, while the embedding for the medium size problem instances (Experiments 1 and 2) are represented in Fig.~\ref{fig:embeddingsExp1Exp2}.

\section{Experimental details}\label{sec:device}
\subsection{The processor chip}

All experiments discussed herein were conducted on a sample
fabricated in a four Nb layer superconducting integrated
circuit process employing a standard Nb/AlOx/Nb trilayer, a TiPt
resistor layer, and planarized SiO$_2$ dielectric layers deposited
with a plasma-enhanced chemical vapour deposition process.  Design
rules included 0.25~$\mu m$ lines and spaces for wiring layers and
a minimum junction diameter of 0.6~$\mu m$.  Experiments  were
conducted in  an Oxford  Instruments Triton 400 Cryofree DR at
a temperature of 20~mK.

\begin{figure}[h]
\includegraphics[width=12cm]{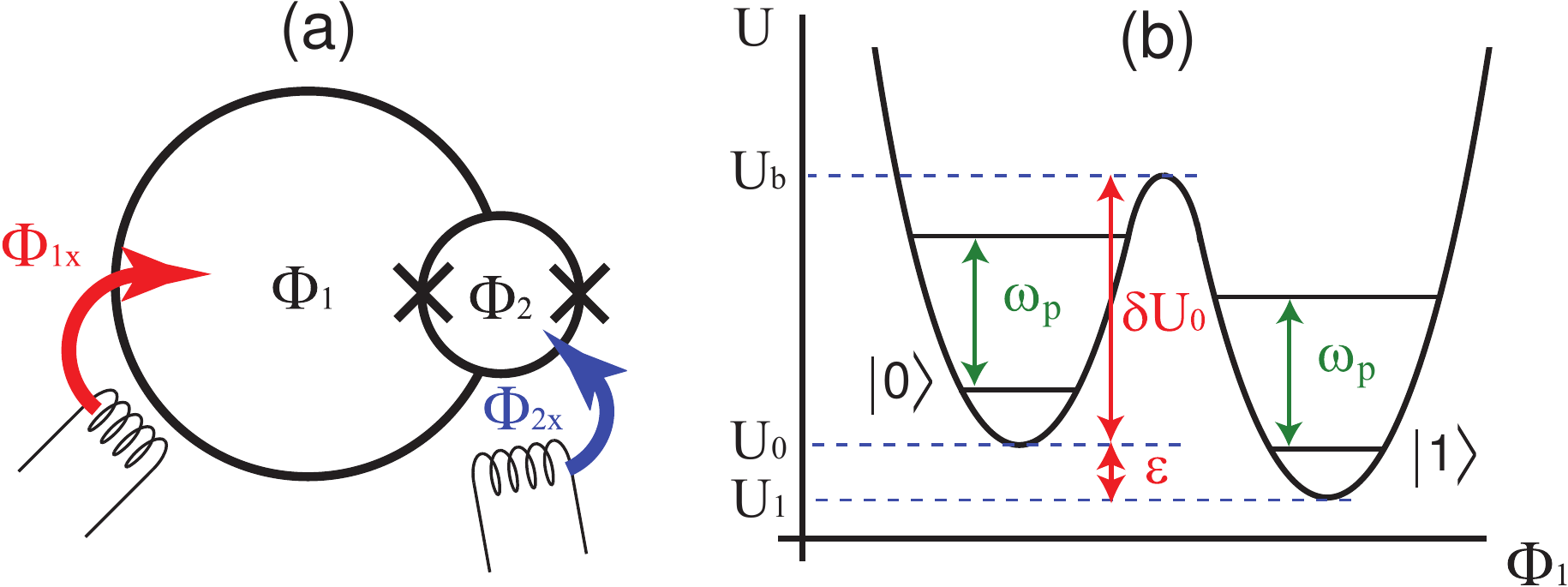}
\caption{ {\bf (a)} Illustration of a single rf-SQUID flux qubit. $\Phi_{1x}$ is the flux bias applied to the major (qubit) loop, and $\Phi_{2x}$ is the flux bias applied to the minor (CJJ) loop. {\bf (b)} Cross-section of the double-well potential of an rf-SQUID flux qubit, with 4 localized energy levels marked. $\Phi_{1x}$ primarily affects the qubit bias $\epsilon$, whereas $\Phi_{2x}$ affects both the barrier height $\delta U_0$ and $\epsilon$.
\label{rfSQUID}}
\end{figure}

The sample processor chip contains a coupled array of 128 qubits
of a design discussed in Ref.~\onlinecite{robust-scalable}. Each
qubit is an rf-SQUID flux qubit with a double-well potential, as depicted in Fig.~\ref{rfSQUID}. They are magnetically coupled with sign
and magnitude tunable couplers in a manner described in
Ref.~\onlinecite{pmm}. The array is built up of 16 eight-qubit unit
cells.  For example, Experiment 6 was conducted using a
single unit cell (highlighted in Fig.~\ref{bipartite}{\bf a}). The
connectivity of qubits within the unit cell is shown schematically
in Fig.~\ref{bipartite}{\bf b}.

Three different chips available with this same architecture were used to run the different problem instances (Experiments 1-6, Fig.~\ref{fig:landscape6AA} and~\ref{fig:landscapeHPPH}). Experiments 1, 2, and 4  were run in one chip, while Experiment 3 and 5 used a different chip. Experiment 6 used the same chip and unit cell used in Ref.~\onlinecite{johnson_quantum_2011}. Since all the chips have the same architecture and design but different calibration parameters, we will focus on the chip used to run Experiment 6, and report all the parameters used to run the numerical simulation reported in Sec.~\ref{sec:qsimulations}.

\subsection{Magnetic Environment}

\begin{figure}[h]
\includegraphics[width=4cm]{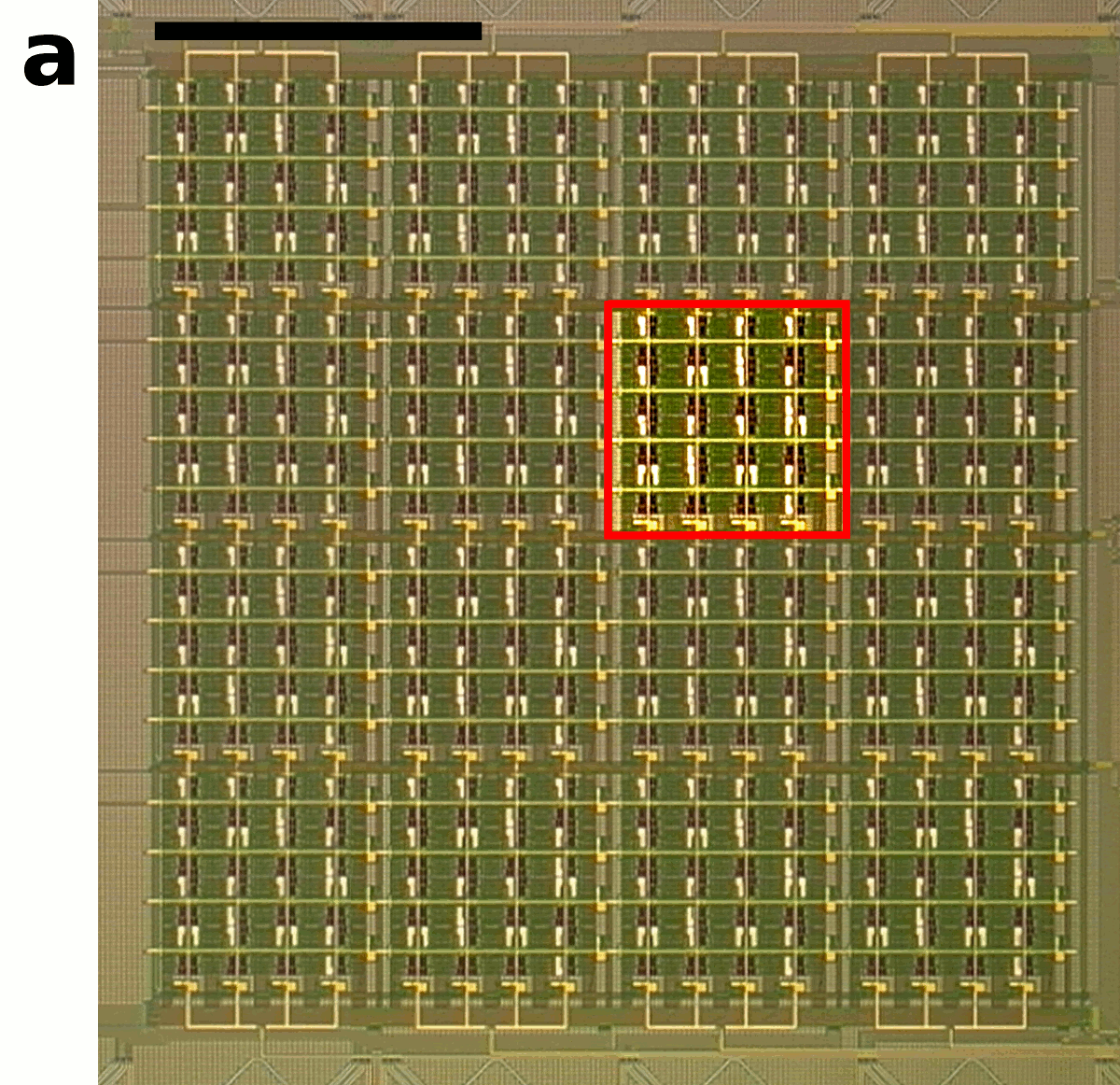}
\hskip1cm
\includegraphics[width=6cm]{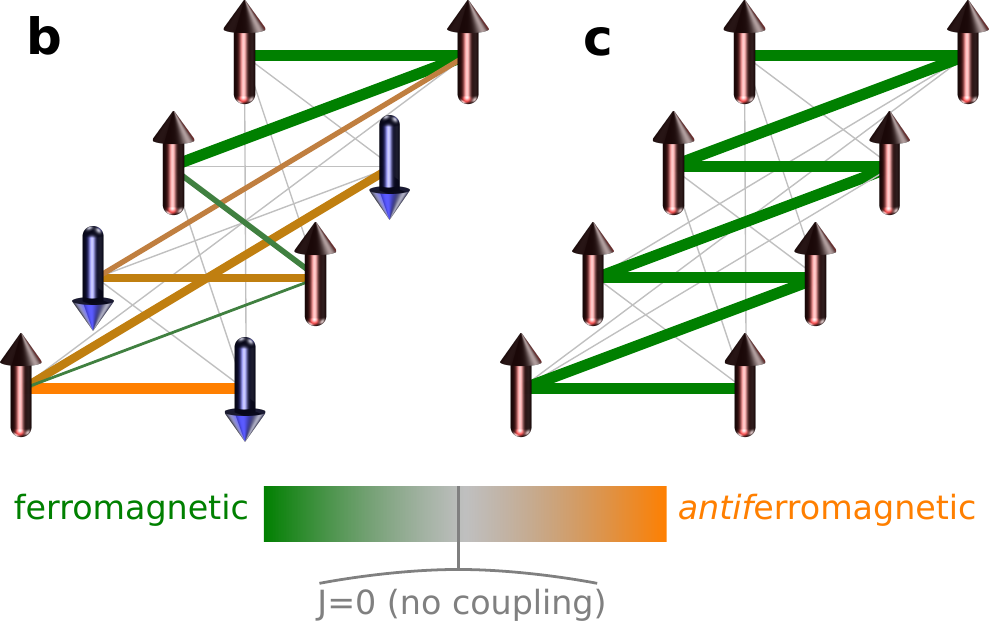}
\caption{ {\bf (a)} Optical photograph of a portion of a partially
fabricated 128 qubit chip. The block of eight qubits used in this
experiment is outlined in red. {\bf (b)} Artificial spins are
connected in a complete bipartite graph $K_{4,4}$, and interact via
couplers which are continuously tunable from ferromagnetic to
antiferromagnetic interaction. A line between artificial spins
indicates that a coupler is present.  The colouring indicates one
possible arrangement of coupler settings. {\bf (c)} An example of
how a linear ferromagnetic Ising spin chain could be implemented by
selectively tuning some couplers to a
ferromagnetic setting $(J < 0)$ (green), and turning off the rest $(J=0)$.
\label{bipartite}}
\end{figure}

The magnetic field in the sample space was controlled with three
concentric high permeability shields and an innermost superconducting
shield.  Further active compensation of residual fields was achieved
with compensation coils oriented along three axes, and used in conjunction
with on-chip superconducting quantum interference device (SQUID)
magnetometers located near each of the four corners of the processor
block (Fig.~\ref{bipartite}{\bf a}).  Compensation coils were adjusted
to minimise the magnetic field measured at the magnetometers while the
chip was at 4.2~K.  The chip was then thermally cycled just above and
then back down through its superconducting transition temperature at
this minimal field.  We estimate that the chip was cooled through its
superconducting transition with a field normal to the chip surface
$\vert B_\perp\vert < 2.5~\mathrm{nT}$, and that parallel to its surface
$\vert B_\parallel\vert < 3.6~\mathrm{nT}$ over the area of active circuitry.

\newpage
\subsection{Experimental method}
\label{section:exp}
The experiment discussed in the manuscript is outlined in Table~\ref{table:experiment}.

\begin{table}[h]
\caption{Outline of experiment \label{table:experiment}}
\begin{enumerate}[{\bf I.}]
\item {\bf Initialisation}
\begin{enumerate}[1.]
\item {\bf Calibration:} measure intrinsic device parameters such as junction $I_c$, qubit inductance, transformer mutual inductances, etc.
\item {\bf Homogenisation:} use on-chip programmable flux biases to ensure $I_p$ of the different qubits match during annealing.
\end{enumerate}
\item {\bf Annealing \& read-out}
\begin{enumerate}[1.]
\item {\bf Set h, J}
\item {\bf Anneal} (reduce $A(\tau)$ and increase $B(\tau)$)
\item {\bf Read} state of spins
\end{enumerate}
\end{enumerate}
\end{table}

The steps in part {\bf I} were performed once and would, in general,
only be performed once for a new chip. The calibration step {\bf I}-1
is performed by measuring the circulating current $I_p$ in each qubit,
and its dependence on the CJJ loop flux bias $\Phi_{2x}$.  From this information, one can extract
the qubit critical current $I_c$ and inductance $L$.
Details of this procedure are discussed in detail in section IV.A of
Ref.~\onlinecite{robust-scalable}.  Given these qubit parameters, the
effective inter-qubit coupling strength attained by the tunable couplers can be
determined. This was done by measuring the difference in magnetic flux
coupled into a qubit B between states $\vert\uparrow\rangle$ and
$\vert\downarrow\rangle$ of a qubit A.  This coupled flux was measured
as a function of the setting of the tunable coupler between qubits
A and B, in a manner described in detail in Ref.~\onlinecite{coupler-paper}.

\begin{figure}[ht]
\includegraphics[width=8cm]{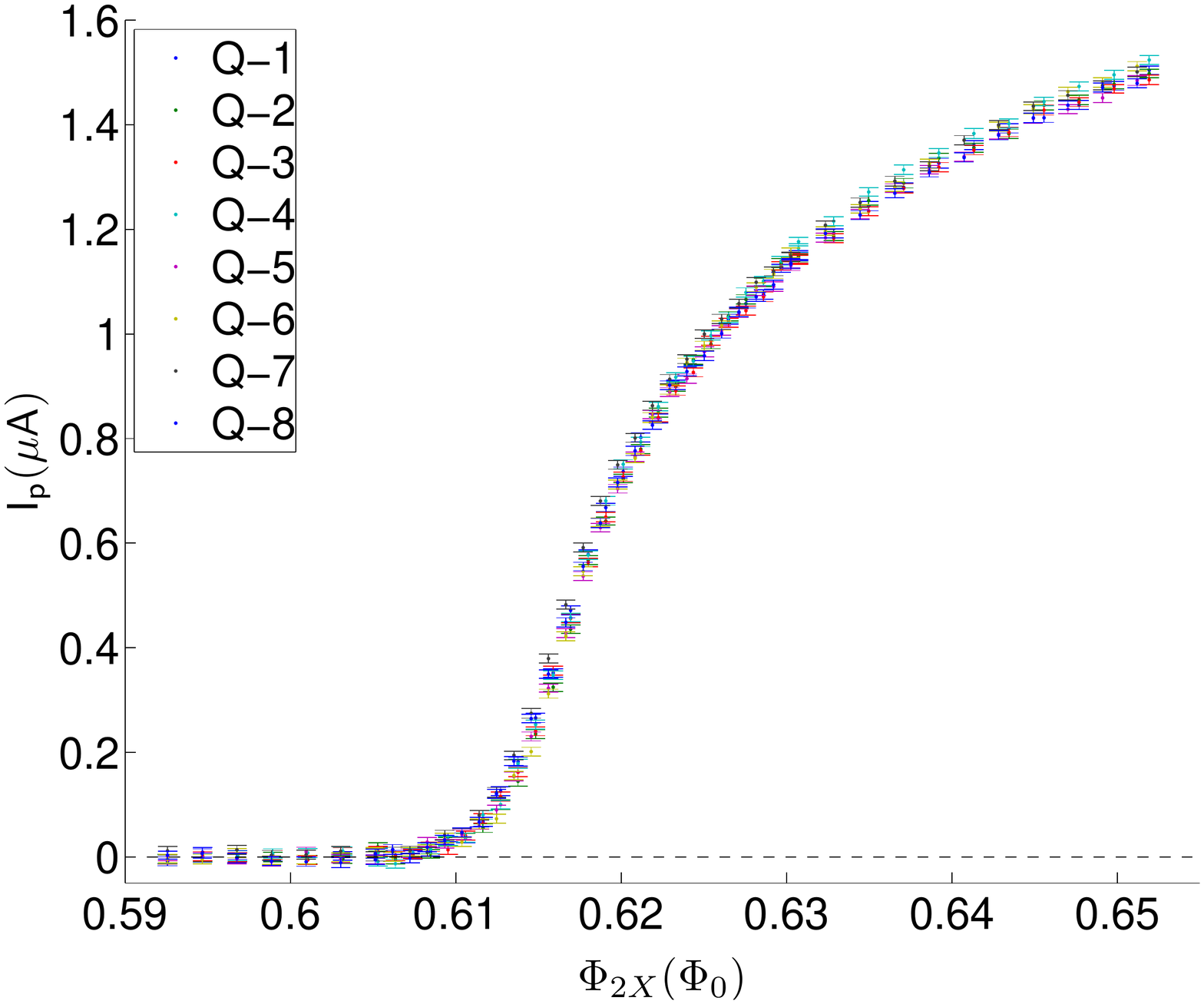}
\includegraphics[width=8cm]{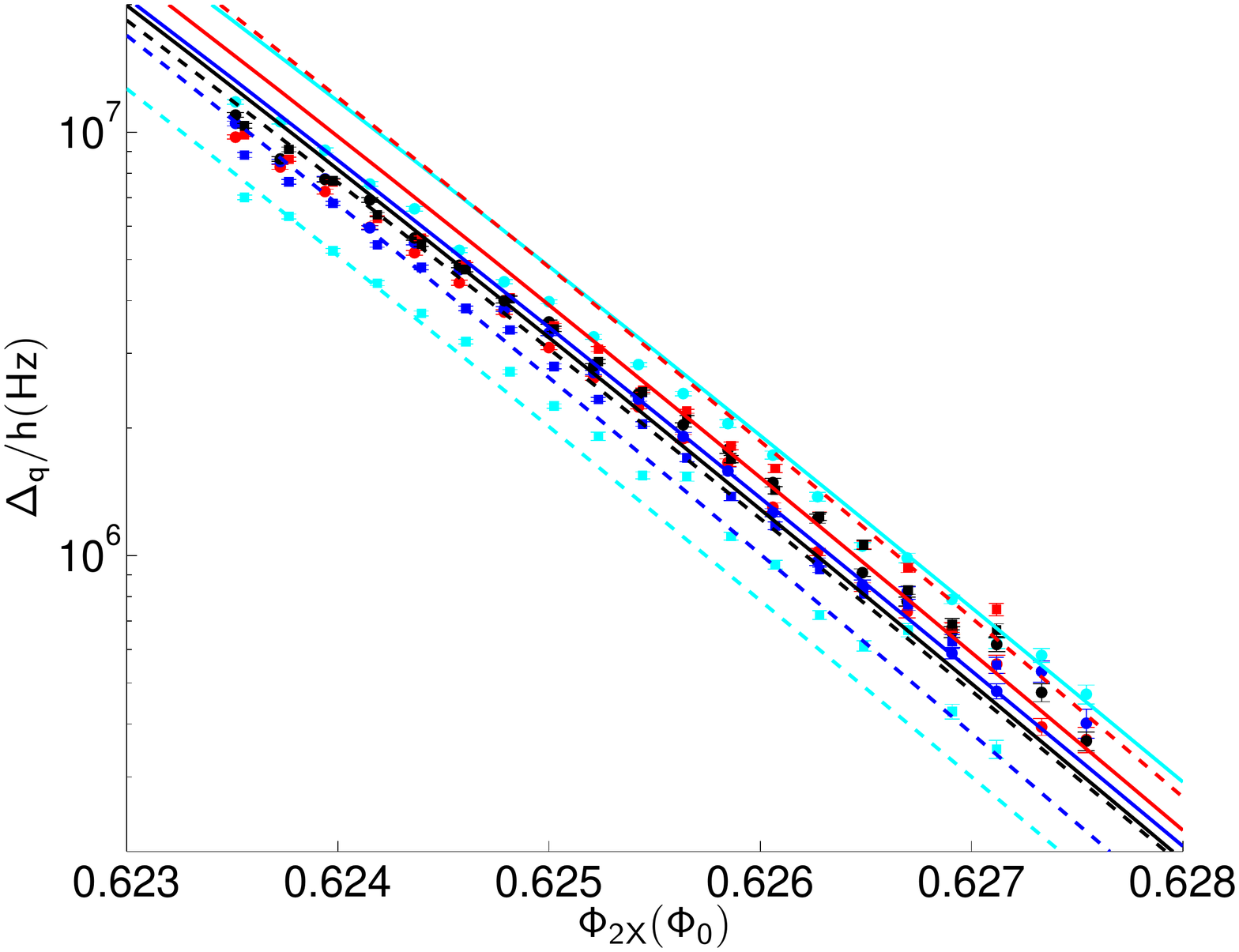}
\caption{(Left) Measured circulating current $I_p$ of each of the
eight qubits used in this experiment after homogenisation
(step {\bf I}-2).
(Right) Comparison of measured tunnel splitting $A(\tau)$ (labelled $\Delta_q$) for the eight qubits used in the experiment, and the $A(\tau)$ fit to a physical model of the rf-SQUID.  Fits of measured tunnel splitting $A(\tau)$ are used in conjunction with  fits to $I_p$ and MRT rate measurements to estimate parameters shown in Table~\ref{qubitparameters}. \label{fig:Ip}}
\end{figure}

Once the device parameters for each qubit have been extracted, the
effective junction $I_c$ and inductance $L$ of each qubit are tuned
with on-chip tuning structures so as to make them as similar to
each other as possible.  The goal of this homogenisation procedure is to ensure that the circulating currents, $I_p$, of several qubits remain close to each other in magnitude
while the qubits undergo annealing. This procedure is discussed in
detail in Refs.~\onlinecite{harris-c1} and \onlinecite{synchronize}.
On-chip tuning structures enabling this homogenisation are also
described in Refs.~\onlinecite{robust-scalable} and \onlinecite{pmm}.
Figure~\ref{fig:Ip} shows the superimposed plots of the measured
circulating current $I_p$ (left) and tunnel splitting $A(\tau)$ (right)
of each of the eight qubits used in this experiment after
homogenisation.  Qubit capacitance is extracted by measuring the
spacing of macroscopic resonant tunnelling rate
peaks\cite{DWaveMRT}.  At any point in $\Phi_{2x}$, the standard
deviation of the measured $I_p$ across the 8 qubits is less than 25 nA.
The uncertainty in each measurement of $I_p$ is about 9~nA.  The
homogenised device parameters are summarised in
Table~\ref{qubitparameters}.

 \begin{table}[h]
 \caption{Total Josephson junction critical current, qubit inductance,
 inductance of loop 2, and junction capacitance
 extracted from circulating current and tunnel splitting measurements,
 and Macroscopic Resonant Tunneling (MRT) peak spacing. \label{qubitparameters}}
 \begin{tabular}{rrrrr}
 Qubit & $I_c(\mu A)$ & $L_1$ (pH) & $L_2$ (pH) & C(fF) \\ \hline
 1 & 3.350 & 337.9 & 26 & 185\\ \hline
 2 & 3.363 & 339.7 & 26 & 190\\ \hline
 3 & 3.340 & 333.0 & 26 & 190 \\ \hline
 4 & 3.363 & 338.5 & 26 & 190 \\ \hline
 5 & 3.340 & 334.0 & 26 & 195 \\ \hline
 6 & 3.352 & 334.8 & 26 & 190 \\ \hline
 7 & 3.365 & 338.8 & 25 & 185 \\ \hline
 8 & 3.330 & 332.9 & 26 & 190 \\ \hline
 \end{tabular}
 \end{table}

The steps in {\bf II} are performed repeatedly.  Step {\bf II}-1 is
where the Hamiltonian parameters $h_i$ and $J_{ij}$ from Eq.~\ref{eq:h_embeddingExp6} are programmed.  For each such problem specification, steps {\bf II}-2 and {\bf II}-3 were
repeated to allow collection of statistics about the relative probabilities of the possible states. For data presented in this paper related to Experiment 6, {\bf II}-1 was repeated 8 times, after each of which, {\bf II}-2 and {\bf II}-3 were repeated 4096 times, for a total of 32,768 repetitions of {\bf II}-2 and {\bf II}-3.  However, step {\bf II}-1 non-negligibly heated the chip, so in order to allow ample time for the chip to cool back to the base temperature, the first 512 repetitions after each execution of step {\bf II}-1 were removed, leaving $8\times 3,584{=}28,672$ total repetitions of {\bf II}-2 and {\bf II}-3. In the case of Experiments 1-5 the statistics were collected over 10,000 measurements in each experiment and enough thermalization time was allowed. Therefore, all data was included in the statistics without the need for removing any of the initial measurements. The experimental results of the probabilities measured are reported as percentages in Figs.~\ref{fig:landscape6AA} and~\ref{fig:landscapeHPPH}.  

\begin{figure}[t]
\includegraphics[width=10cm]{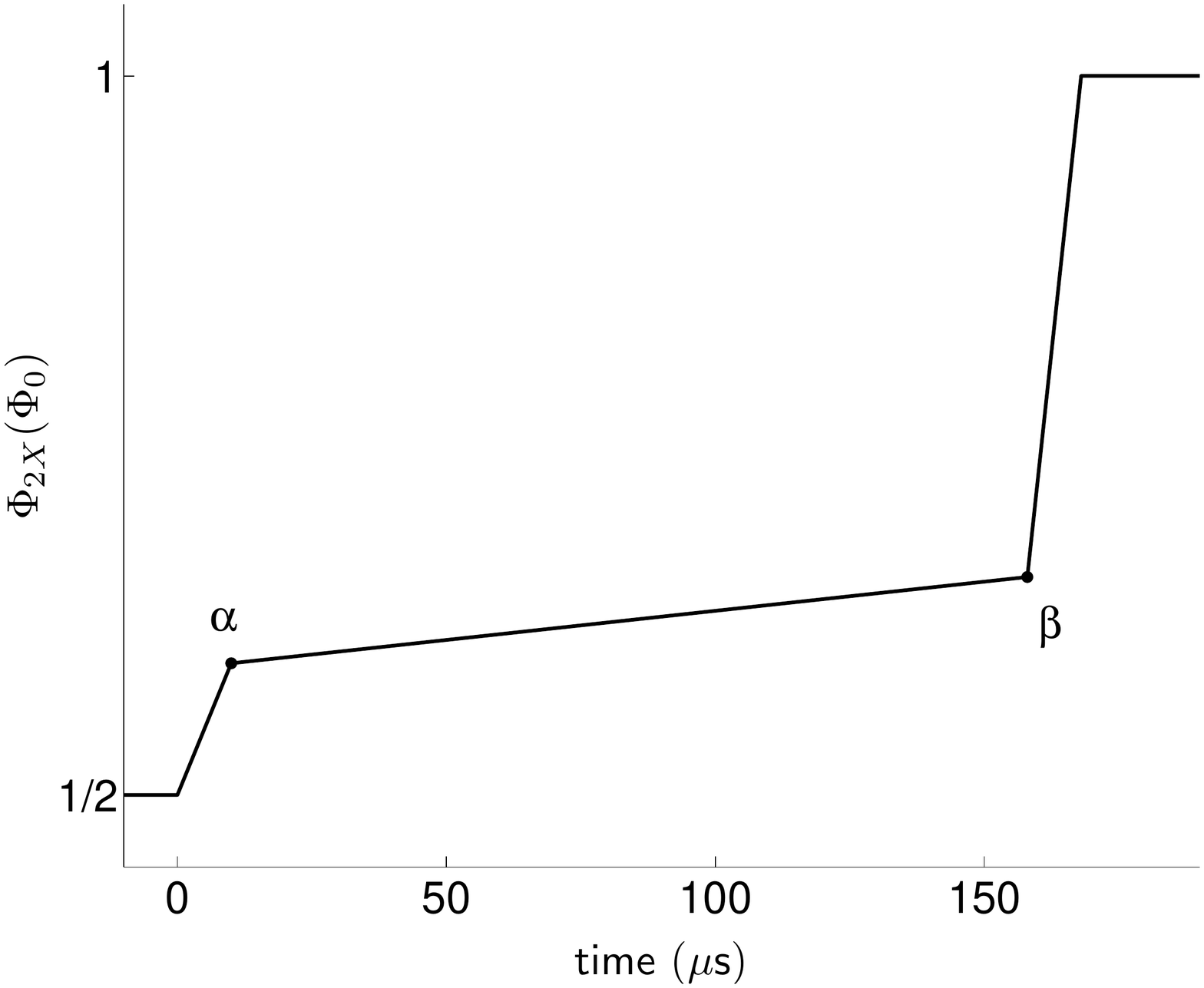}
\caption{The annealing schedule is defined by the applied flux
$\Phi_{2x}(t)$.  The qubits make a transition between being monostable
and bistable between control points $\alpha = (10~\mu\mathrm{s},-0.592~\Phi_0)$
and $\beta = (158~\mu\mathrm{s},-0.652~\Phi_0)$.  \label{annealingschedule}}
\end{figure}

Annealing was performed by raising the single qubit tunneling barrier.  This is accomplished by changing $\Phi_{2x}$ linearly in time, from $0.592~\Phi_0$
to $0.652~\Phi_0$, over a period of $148~\mu\mathsf{s}$, as shown in
Figure~\ref{annealingschedule}.  Circulating current $I_p$ shown in
Figure~\ref{fig:Ip} is plotted over exactly this range of $\Phi_{2x}$.
This also has the effect of changing parameters $A(\tau)$ and $B(\tau)$
from Eq.~(3) of the main paper, as shown in Fig.~1{\bf b} of the
main paper, and as discussed in Ref.~\onlinecite{controlled-anneal}.
Control points $\alpha$ and $\beta$ in Figure~\ref{annealingschedule}
correspond to the beginning and ending times of Fig.~2(b) of the main text.

After the qubits have completed annealing, when $\Phi_{2x}$ has been
set to $\Phi_0$ as shown on the right in
Figure~\ref{annealingschedule}, states of the spins are read with
a hysteretic dc-SQUID readout, as described in
Ref.~\onlinecite{readout}.

\subsection{Thermometry}

In addition to a Ruthenium Oxide thermometer mounted on the dilution refrigerator mixing chamber, the effective qubit device temperature obtained during the measurements discussed in the manuscript was determined in two independent ways.  The first is based on analysis of the single-qubit Macroscopic Resonant Tunnelling (MRT) rate, and its dependence on the qubit loop flux bias $\Phi_{1x}$.  Measurements and analysis of MRT rates for the devices used in this experiment are discussed in Ref~\onlinecite{Lanting2011}.  The second is based on measurement of the equilibrium $P_\uparrow$ vs. $\Phi_{1x}$ attained at fixed barrier height (fixed value of $\Phi_{2x}$).  Both of these techniques are discussed in some detail in Ref.~\onlinecite{DWaveMRT}.

At a fixed barrier height achieved with a fixed value of $\Phi_{2x}$, the equilibrium probability $P_\uparrow$ approaches the thermal distribution:
\be P_{\uparrow}(t \rightarrow \infty) = \frac{1}{2}\left[\frac{1}{2} + \tanh\left(\frac{I_p \Phi_{1x}}{k_B T_{th}}\right)\right]
\label{eq:tth}
\ee
where $I_p$ is the value of circulating current obtained at that value
of $\Phi_{2x}$ and  $T_{th}$  is the  effective device  temperature.
Fitting a measurement of $P_\uparrow$  as a function of $\Phi_{1x}$ to
Eq.~\ref{eq:tth}, combined with a knowledge of $I_p$, allows us to
extract $T_{th}$.

\begin{figure}[h]
\includegraphics[width=8cm,angle=270]{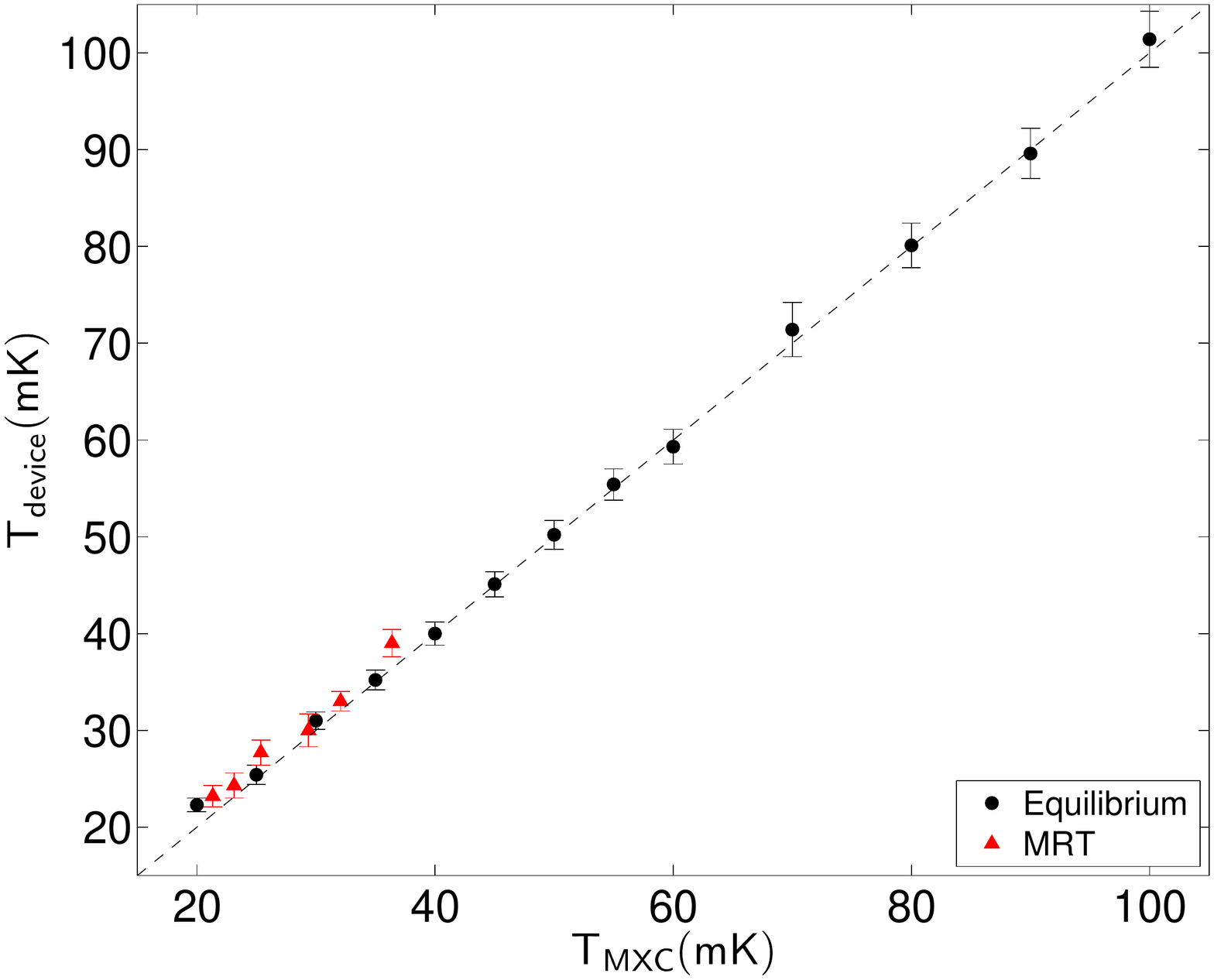}
\caption{Plots of $T_{th}$ (black circles) and $T_{MRT}$ (red triangles)
vs. the temperature measured with the Ruthenium Oxide thermometer mounted
on the mixing chamber $T_{MXC}$.\label{figure:thermometry}}
\end{figure}

Measurement of  $T_{th}$ was performed on  two of the  devices at each
temperature  setting.    An  average  of  at   least  two  independent
measurements of the device temperature  $T_{th}$ of each of two qubits
is compared against the mixing chamber thermometer temperature reading
($T_{MXC}$)   in   Figure~\ref{figure:thermometry}.   Uncertainty   in
$T_{th}$ was dominated by the  uncertainty in the fit transition width
for each measurement, which was  generally found to be larger than the
standard deviation of the separate measurements.

The temperature extracted from MRT transition rate widths ($T_{MRT}$) is
also  plotted   vs.  $T_{MXC}$   for  temperatures  below   40~mK,  in
Figure~\ref{figure:thermometry}.   From these plots  it is  clear that
the two  methods generally  agree with each  other as well as with the mixing
chamber thermometer to within 3~mK over the temperature range used
in the experiment.

\section{Quantum simulations}\label{sec:qsimulations}

To obtain better quantitative understanding of the behaviour of the system, a simulation was conducted to model this experiment. The agreement between the numerical simulations can be seen in panel (b) of Fig.~\ref{fig:landscapeHPPH}, where both percentages [experiment (theory)] are reported next to each other for each one of the low-energy conformations.

 Our simulation strategy is as follows: We first write a Hamiltonian for the superconducting circuit based on standard circuit models for capacitances, inductances, and Josephson junctions. This Hamiltonian is expected to correctly describe the behaviour of coupled rf-SQUIDs. We then numerically calculate the evolution of the system based on this Hamiltonian using quantum mechanical equations of motion which take into account coupling to an environment.  Therefore, we predict the quantum evolutions for the same system Hamiltonian, the same coupling to environment, and the same type of noise spectral densities. This provides a fair comparison to the experimental data.

\subsection{rf-SQUID Hamiltonian}

A simplified version of the rf-SQUID qubit used in our processor is illustrated in Fig.~\ref{rfSQUID}{\bf a}.  (A more complete description of the actual qubits can be found in Ref.~\onlinecite{robust-scalable}.) It has two main superconducting loops and therefore two flux degrees of freedom $\Phi_1$ and $\Phi_2$, subject to external flux biases $\Phi_{1x}$ and $\Phi_{2x}$, respectively. The Hamiltonian of such an rf-SQUID is written as \ba H_{\rm SQUID} = {q_1^2 \over 2C_1} + {q_2^2 \over 2C_2} + U(\Phi_1,\Phi_2) \label{HS} \ea where $C_1$ and $C_2$ are parallel and series combinations of the junction capacitances, $q_1$ and $q_2$ are the sum and difference  of the charges stored in the two Josephson junctions respectively, and \ba U(\Phi_1,\Phi_2) = {(\Phi_1-\Phi_{1x})^2/2L_1} + {(\Phi_2-\Phi_{2x})^2/2L_2} \nn - 2E_J \cos(\pi \Phi_2/\Phi_0)\cos(2\pi \Phi_1/\Phi_0), \qquad \label{U} \ea is a 2-dimensional potential with $L_i$ being the inductances of the two loops and $\Phi_0 {=}\, h/2e$, the flux quantum.  We have assumed symmetric Josephson junctions with Josephson energies $E_J {=}\, I_c\Phi_0/2\pi$, where $I_c$ is the junctions' critical current. (A small asymmetry can be tuned away in situ in the physical implementation \cite{robust-scalable}.)

At $\Phi_{1x}\approx \Phi_0/2$, the potential can become bistable
and therefore form a two-dimensional double-well potential. If $L_2$
is small enough so that the deviation of $\Phi_2$ from $\Phi_{2x}$
can be neglected, then the two-dimensional classical potential
$U(\Phi_1,\Phi_2)$ can be approximated by a one-dimensional
double-well potential, as shown in Fig.~\ref{rfSQUID}{\bf b}.
However, with our realistic qubit parameters, $\Phi_2$ cannot be
neglected and therefore is accounted for in all our numerical
calculations. When $\Phi_{1x} = \Phi_0/2$, the two wells are
symmetric with no energy bias between them. One can tilt the
potential by changing $\Phi_{1x}$ and establish an energy bias, as
depicted in Fig.~\ref{rfSQUID}{\bf b}. It is also possible to change
the barrier height by changing $\Phi_{2x}$.

An array of such qubits can be modelled by summing contributions
of Eq.~(\ref{HS}) from each device plus terms that describe
magnetic coupling of the loops:
 \be
 H_S = \sum_i H_{\rm SQUID}^{(i)} + \sum_{i>j} H^{(ij)}_{\rm coupl}
 \label{Hsystem}
 \ee Coupling between  qubits $i$ and $j$ can be  modelled as a mutual
 inductance $M_{ij}$  between loop 1  of each pair of  coupled qubits:
 \be H^{(ij)}_{\rm coupl} =
 (\Phi^{(i)}_1-\Phi^{(i)}_{1x})(\Phi^{(j)}_1-\Phi^{(j)}_{1x})M_{ij}/L_1^{(i)}L_1^{(j)}
 \ee
As discussed in Section~\ref{section:exp} above, all parameters,
i.e., inductances $L_\alpha^{(i)}$, capacitances $C_\alpha^{(i)}$,
and Josephson critical currents $I_{c}^{(i)}$, are measured
independently for each qubit.

To describe the system accurately we also need to introduce
interaction with environment. Flux noise, which is the dominant
noise in flux qubits, couples to the $i$th qubit as fluctuations
$\delta \Phi_{\alpha x}^{(i)}$ of the external flux $\Phi_{\alpha
x}^{(i)}$:
 \be
 H_{\rm int} = -\sum_{\alpha = 1}^2 \sum_i
 {\Phi_\alpha^{(i)} - \Phi_{\alpha x}^{(i)} \over L_\alpha^{(i)}} \delta\Phi_{\alpha
 x}^{(i)}
 \label{Hint}
 \ee
The noise is much smaller for the smaller loop $\Phi_{2x}^{(i)}$
than for the larger loop $\Phi_{1x}^{(i)}$ due to the loop size. The
flux noise $\delta \Phi_{\alpha x}^{(i)}$ is assumed to be
uncorrelated between the qubits, which agrees with recent
experimental observation\cite{PhysRevB.82.060512}.

\subsubsection{Chip calibration and device parameter extraction}

Device parameters were extracted for the simulations through a
series of  independent  measurements  of  qubit circulating current,
tunnel splitting $\Delta$, and  MRT peak spacing.  A discussion  of
how these measurements        are       performed is        given in
Ref.~\onlinecite{robust-scalable}. Parameter values used in simulations
are summarised in Table~\ref{qubitparameters}.

\subsection{Quantum Simulation}

To simulate the quantum mechanical dynamics of the system, we treat (\ref{Hsystem})-(\ref{Hint}) as quantum mechanical Hamiltonians. In that case, the charge $q^{(i)}_\alpha$ is taken to be an operator, which is the momentum conjugate to the flux operator $\Phi^{(i)}_\alpha$ with commutation relation: $[\Phi^{(i)}_\alpha,q^{(i)}_\alpha]=i\hbar$. Unfortunately, it is impossible to calculate the dynamics of the system directly on the $2N$-dimensional continuous potential quantum mechanically. Instead, we use energy discretization as a means to simplify the calculation. The simplest way to accomplish this is to treat an rf-SQUID as a 2-state system or qubit and replace (\ref{Hsystem}) by a coupled qubit Hamiltonian. One may go further and keep more than two states per rf-SQUID in the calculation, as we shall discuss below.

We first numerically diagonalise the single rf-SQUID Hamiltonian
(\ref{HS}) to obtain the lowest eigenvalues and eigenvectors. We treat
the lowest few energy levels as the subspace relevant for
computation. We then write the Hamiltonian in the basis of states
that are localised within the wells. Such states are not true
eigenfunctions of the Hamiltonian and, therefore, are metastable
towards tunnelling to the opposite well. Hence, the resulting
Hamiltonian in such a basis will have off-diagonal terms between
states in the opposite wells but not between states within each
well. The latter is because those states should be stationary within
their own wells; any transition (relaxation) between them is only
induced by the environment.

Let $|l\rangle$ denote localised states within the wells. We use
even (odd) state numbers, i.e., $l=2n$ ($2n{+}1$), with
$n=0,1,2,...$, to denote states that are localised in the left
(right) well. For the lowest $M$ energy levels ($M$ is taken to be
even), the effective $M {\times} M$ tunnelling Hamiltonian is written
as
 \ba
 H_S = \sum_{l=0}^{M-1} E_l|l\rangle\langle l| +
 \sum_{n,m=0}^{M/2-1} K_{2n,2m{+}1}
 (|2n\rangle\langle 2m{+}1|+|2m{+}1\rangle\langle 2n|) \label{Htunneling}
 \ea
where $E_l$ is the energy expectation value for state $|l\rangle$
and $K_{2n,2m{+}1}$ is the tunnelling amplitude between states
$|2n\rangle$ and $|2m{+}1\rangle$, which exist in opposite wells.
Notice that there is no matrix element between states on the same
well: $\langle 2n|H_S|2m\rangle = \langle 2n{+}1|H_S|2m{+}1\rangle =
0$, which means that the states are metastable only towards
tunnelling to the other side, or the states are quasi-eigenstates of
the Hamiltonian within their own sides. All parameters of the
tunnelling Hamiltonian, i.e., $E_l$ and $K_{ll'}$ are extracted from
the original rf-SQUID Hamiltonian (\ref{HS}). For the 2-state qubit
model we keep only the lowest two energy levels of
(\ref{Htunneling}). The effective qubit Hamiltonian can be written
as
 \ba
 && H_{eff} = -{1\over 2}(\epsilon\sigma_z + \Delta\sigma_x)
 \ea
where
 \ba
 \epsilon = E_0 - E_1, \qquad
 \Delta = -2K_{01},
 \ea

We also go beyond the 2-state model and keep 4 states per rf-SQUID. Those 4 states can be
represented by two coupled qubits, one of which represents the direction of
persistent current or flux, and the other one generating intrawell energy levels. We
represent the first (logical) qubit by Pauli matrices $\sigma_\alpha$, and the extra
(ancilla) qubit by Pauli matrices $\tau_\alpha$. The effective Hamiltonian for those
two coupled qubits can be written as
 \ba
 H_{eff} = -{1\over 2}(\epsilon\sigma_z + \Delta\sigma_x)
 + {1\over 2}[\omega_p\tau_z + \kappa_{xz} \sigma_x(1+\tau_z) + \kappa_{xx}
 \sigma_x\tau_x ]. \label{Hqudit}
 \ea
It is easy to show that (\ref{Hqudit}) is equivalent, up to a
constant energy, to Hamiltonian (\ref{Htunneling}), with $M=4$,
if
 \be
 \epsilon = E_0 - E_1 = E_2 - E_3, \qquad
 \omega_p = E_2 - E_0 = E_3 - E_1, \qquad
 \Delta = -2K_{01},
 \ee
 \be
 \kappa_{xz} = K_{23}-K_{01} \approx K_{23}, \qquad
 \kappa_{xx} = 2K_{03} = 2K_{12}.
 \ee
As can be seen, the coupling between logical and ancilla qubits are
of XX+XZ type. Coupling qubits to each other is accomplished using $\sigma_z$
operators which represent the direction of the induced flux. The
ancilla qubits remain uncoupled from each other and from other
qubits. It should be noted that the readout at the end of the
evolution can only distinguish ``left" well from ``right" well in
the double-well potential and cannot distinguish levels within each
well. This is equivalent to reading out logical qubits and not
ancilla qubits, but as we mentioned above, only logical qubits carry information.

\begin{figure}[t]
\includegraphics[width=7cm]{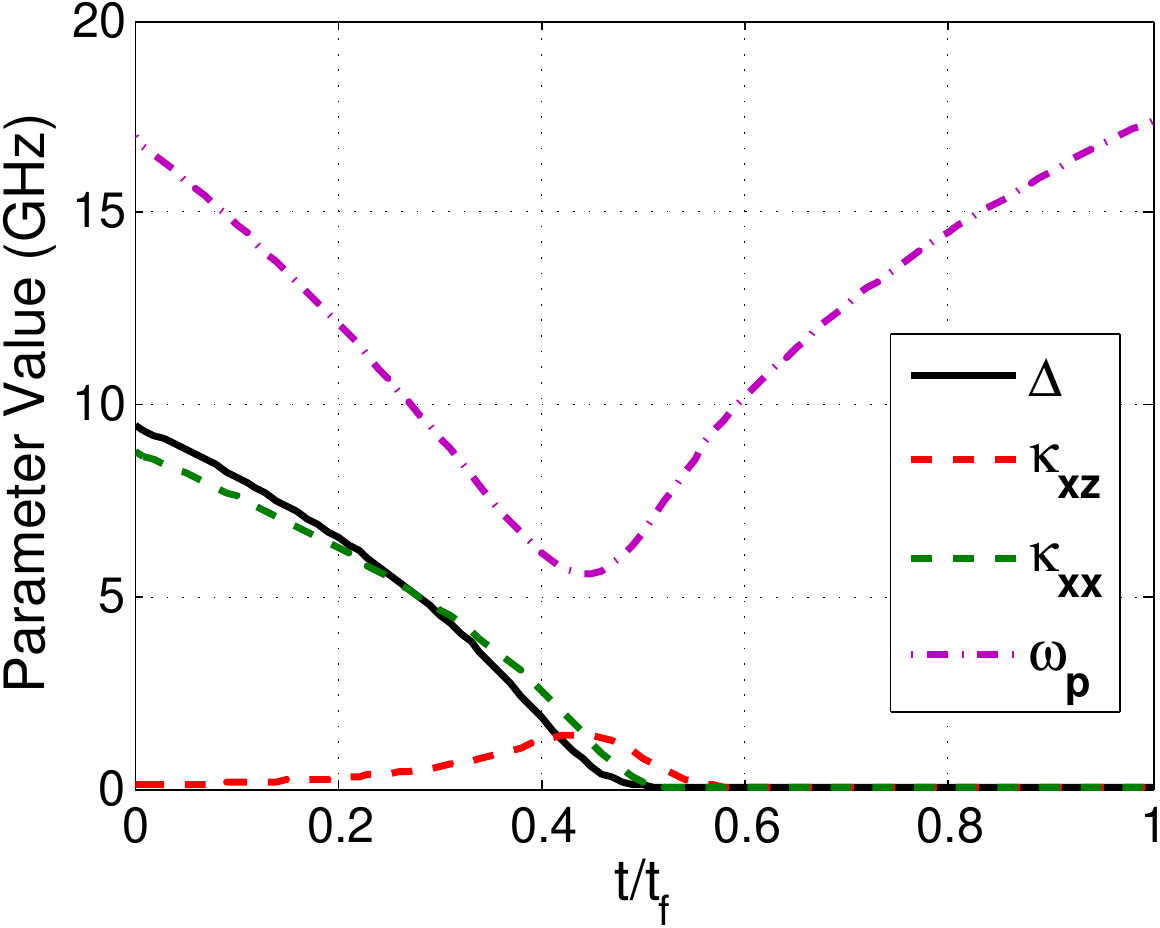}
\caption{Parameters of the 4-level model for rf-SQUID qubit as a
function of time during annealing. \label{parameters}}
\end{figure}

To properly treat the environment, we need to write the interaction
Hamiltonian (\ref{Hint}) in the subspace of the lowest energy levels
in terms of Pauli matrices. For quantum simulations we only consider
noise coupling to the larger loop in Fig.~\ref{rfSQUID}{\bf a}. Let us consider a single rf-SQUID and write the interaction
Hamiltonian as
 \be
 H_{\rm int} = -{\Phi_1 - \Phi_{1x} \over L_1} \delta\Phi_{1x}
 \label{Hint_rfSQUID}
 \ee
We define the qubit persistent current by
 \be
 I_p =  {1\over L_1}\left|\langle l|(\Phi_1-\Phi_{1x})|l\rangle\right|.
 \ee
Here, we take $I_p$ to be independent of $|l\rangle$ for the low
lying states considered, although in reality there could be a small
dependence. The interaction Hamiltonian can then be written as
 \be
 H_{\rm int} = -{1 \over 2}(\sigma_z + \lambda \tau_x) Q, \label{Hint0}
 \ee
where
 \be
 Q=2I_p \delta \Phi_{1x}, \qquad
 \lambda = {\langle 0|H_{\rm int}|2\rangle \over 2 I_p L_1}
 = {\langle 1|H_{\rm int}|3\rangle \over 2 I_p L_1}
 \ee
The matrix elements $\langle 0|H_{\rm int}|2\rangle$ or $\langle
1|H_{\rm int}|3\rangle$ are calculated directly via Eq.~\ref{Hint_rfSQUID} using the eigenfunctions of the rf-SQUID Hamiltonian, Eq.~\ref{HS}). The values of $I_p$ and $\lambda$ can therefore be calculated numerically from the original rf-SQUID Hamiltonian. Only $Q$ remains  which should be
characterised via its spectral density, which is the subject of Appendix A.

Quantum evolution of the system was calculated using a Markovian
master equation for the density matrix described in Ref.~\onlinecite{ATA09}.
Since the evolution is very slow (adiabatic) and temperature is low,
only a small number of energy levels are expected to be occupied
during the evolution. We write the density matrix in the
instantaneous energy eigenstate basis and truncate it to the lowest
24 energy levels, which was found to sufficiently describe the type of evolution studied here. We use both 2-state and
4-state models for rf-SQUIDs, as described above, in our
simulations. The result of the 4-level model simulation is shown in
Fig.~\ref{fig:landscapeHPPH}{\bf b}.

\appendix
\section{Noise spectral density}

The quantum noise operator $Q=2I_p \delta \Phi_{1x}$ is related to the
flux noise as expected (for simplicity we only consider one
rf-SQUID), and is characterised by its correlation function. Let us define the spectral density
 \be
 S(\omega) = \int_{-\infty}^\infty dt \ e^{i\omega t}
 \langle Q(t)Q(0) \rangle = 4I_p^2 S_\Phi(\omega)
 \ee
where
 \be
 S_\Phi(\omega) = \int_{-\infty}^\infty dt \ e^{i\omega t}
 \langle \delta\Phi_{1x}(t)\delta\Phi_{1x}(0) \rangle
 \ee
is the spectral density of the flux noise. No direct
measurement of $S_\Phi(\omega)$ at all frequencies is available. We
assume the spectral density is a sum of low frequency and high
frequency components: $S_\Phi(\omega) =
S_{LF}(\omega)+S_{HF}(\omega)$. For the low frequency component we
use
 \be
 S_{LF}(\omega) =  {(A^2/k_BT) \hbar\omega|\omega|^{-\alpha}
 \over 1 - e^{-\hbar\omega/k_BT}},
 \ee
with $\alpha \approx 1$, which at low $\omega$ behaves as 1/f noise:
$\sim A^2|\omega|^{-\alpha}$.  Parameter $A$ is measured from low
frequency noise measurement \cite{trevornoise} and is found to be $A
\approx 3$ n$\Phi_0$.

The high frequency parts of the spectral density is assumed to
be ohmic,
 \be
 S_{HF}(\omega) = \left({\hbar^2 \over 4I_{p0}^2}\right)
 {\eta \omega e^{-|\omega|/\omega_c}
 \over 1 - e^{-\hbar\omega/k_BT}},
 \ee
where $\omega_c$ is the upper cutoff frequency, $\eta$ is the
dimensionless coupling coefficient, and $I_{p0}$ is the value of
persistent current at which $\eta$ is measured. The coupling
coefficient and the persistent current are found, using Macroscopic
resonant tunnelling experiment (MRT), to be $\eta \approx 0.4$ and
$I_{p0}\approx 1\ \mu$A. The details of extraction of $\eta$ via MRT
are presented elsewhere~\cite{Lanting2011}.

This leaves no free parameters for the quantum simulations.

\end{document}